\documentclass[12pt,nofootinbib]{article}
\pdfoutput=1
\usepackage{amsmath,amssymb,graphicx,multicol} 
\usepackage{epsf,color}
\usepackage[nosort]{cite}

\ifx\pdfoutput\undefined
\usepackage[dvips,bookmarks=false]{hyperref}	
\else
\usepackage{hyperref}	
\fi

\newcommand{\eps}{\epsilon}
\newcommand{\Tr}{\text{Tr}}

\begin{document}
\begin{titlepage}
\begin{flushright}
CERN-PH-TH-2015-280
\end{flushright}
\vskip1.3cm

\begin{center}
{\Large \bf 
Dispersion Relations for Electroweak Observables \\[0.15cm] in Composite Higgs Models
}
\end{center}
\vskip0.5cm

\renewcommand{\thefootnote}{\fnsymbol{footnote}}
\begin{center}
{\large  Roberto Contino$\,^{1,2}$\footnote{\hspace{0.17cm}On leave of absence from Universit\`a
di Roma La Sapienza and INFN, Roma, Italy.} and Matteo Salvarezza$\,^3$
}
\end{center}
\renewcommand{\thefootnote}{\arabic{footnote}}
\setcounter{footnote}{0}

\vspace{0.1cm}
\begin{center}
{\it 
$^1\,$Institut de Th\'eorie des Ph\'enomenes Physiques, EPFL, Lausanne, Switzerland \\
$^2\,$Theory Division, CERN, Geneva, Switzerland \\
$^3\,$Dipartimento di Fisica, Universit\`a di Roma ``La Sapienza'' and INFN, Roma, Italy
} \\
\vspace*{0.1cm}
\end{center}

\vglue 1.0truecm

\begin{abstract}
\noindent 
We derive dispersion relations for the  electroweak oblique observables measured at LEP
in the context of $SO(5)/SO(4)$ composite Higgs models.
It is shown how these relations can be used and must be modified when modeling the spectral functions through a low-energy 
effective description of the strong dynamics.  The dispersion relation for the parameter $\eps_3$ is then used to estimate the contribution
from spin-1 resonances at the 1-loop level. Finally, it is shown that the sign of the contribution to the $\hat S$ parameter 
from the lowest-lying spin-1 states is not necessarily positive definite, but depends on 
the energy scale at which the asymptotic behavior of current correlators is attained.
\end{abstract}

\end{titlepage}

\section{Introduction}
\label{sec:Intro} 

\enlargethispage{0.5cm}
Theories with strong electroweak symmetry breaking are severely constrained by the electroweak precision observables measured at LEP, SLC and Tevatron.
Large corrections to vector boson polarizations, especially those encoded by the Peskin-Takeuchi $S$ parameter~\cite{Peskin:1991sw}, were  the 
most severe problem of Technicolor theories~\cite{TC}, together with flavor, before the discovery of a light Higgs boson.
To date, electroweak  tests set the strongest constraints on composite Higgs theories~\cite{Kaplan:1983fs,compositeHiggs}, 
and this is even more true for their  recent Twin Higgs realizations~\cite{Chacko:2005pe,Batra:2008jy,Geller:2014kta,Barbieri:2015lqa,Low:2015nqa}.
However, while corrections to electroweak observables can be naively estimated to be generally large, their precise determination in the context of strongly-interacting
dynamics is a challenge. A first-principle approach based on a non-perturbative method such as lattice gauge theories 
is possible but demanding in terms of theoretical efforts and computational power (see for example Refs.~\cite{Sonthelattice} for  calculations of the $S$ parameter 
on the lattice). Simpler, though less rigorous approaches include a variety of perturbative methods like the inclusion of chiral logarithms, effective models of the 
lowest-lying resonances, and the large-$N$ expansion.  Especially powerful in this sense is the 5-dimensional perturbative approach of holographic theories, which 
allows one to effectively resum the corrections of a whole tower of states, the Kaluza-Klein excitations, neglecting smaller effects from string modes.

An alternative strategy consists in making use of dispersion relations to express an observable as the integral over the spectral functions of the strong dynamics.
Extracting the spectral functions from experimental data thus leads to a result which is, at least in principle,
free from theoretical ambiguities. The most successful application of this idea is perhaps the determination of the correction from the electromagnetic vacuum polarization
due to QCD to the muon $g-2$~\cite{gm2}, though equally famous is the estimate of the $S$ parameter in Technicolor theories made by Peskin and
Takeuchi in their seminal paper~\cite{Peskin:1991sw} (where they also compute  the chiral coefficient~$l_{5}$ using the dispersive formula
first derived by Gasser and Leutwyler~\cite{Gasser:1983yg}).
Although the most powerful use of dispersion relations is in conjunction with experimental data, in the absence of the latter one can make
models of the spectral functions based on theoretical considerations. Computing the spectral functions through a low-energy effective theory of resonances
leads in fact to the same result obtained by a more conventional diagrammatic technique, though  
the dispersive approach can simplify the calculation and gives a different viewpoint.

The first application of dispersion relations to composite Higgs theories was given in Ref.~\cite{Orgogozo:2012ct} by Rychkov and Orgogozo, who derived a dispersion
formula for the parameter $\eps_3$ defined by Altarelli and Barbieri~\cite{Altarelli:1990zd}.
A dispersive 1-loop calculation of the $S$ parameter was later performed by Ref.~\cite{Pich:2013fea} (see Appendix B therein).
The aim of this paper is to give an alternative derivation and extend the work of Ref.~\cite{Orgogozo:2012ct} by
obtaining spectral representations for the electroweak parameters $\hat S$, $W$ and $Y$ of Ref.~\cite{Barbieri:2004qk}.
We will focus on $SO(5)/SO(4)$ models as simple though representative examples of composite Higgs theories; the extension to other cosets is straightforward.
We will then use the dispersion formula for $\eps_3$ to estimate the contribution from spin-1 resonances at $O(m_W^2/16\pi^2 f^2)$ by computing 
the  spectral functions in a low-energy effective theory. The result will be shown to coincide with the one we obtained in Ref.~\cite{Contino:2015mha} through a 
diagrammatic calculation. The different viewpoint offered by the dispersive approach will allow us to clarify an issue on the positivity of $\hat S$ raised
in Ref.~\cite{Orgogozo:2012ct}.

The paper is organized as follows. in Section~\ref{sec:dispersionrelation} we review the definition of $\eps_3$ by distinguishing between long- and short-distance 
contributions. Short-distance contributions, in particular, will be parametrized in terms of $\hat S$, $W$, $Y$  and $X$. We derive expressions for $\hat S$, $W$ and $Y$
in terms of two-point current correlators of the strong dynamics, which can be used for a non-perturbative computation on the lattice.
Section~\ref{subsec:dispersionforshortdistance} contains a derivation of the dispersion relation for $\hat S$, $W$ and $Y$, extending the work of Peskin and Takeuchi 
to the case of $SO(5)/SO(4)$ theories. A dispersive formula for $\eps_3$ is then derived. The result is shown to agree with the previous result of Rychkov and Orgogozo, 
and improves on it by reducing the relative uncertainty.
In Section~\ref{sec:disersiveEFT} we show how dispersion relations can be used and must be modified in order to model the spectral functions in the context of a low-energy
effective description of the strong dynamics.  The dispersion relation for $\eps_3$ is then used in Section~\ref{sec:calculation} to estimate the contribution
from spin-1 resonances at the 1-loop level. We discuss the positivity of $\hat S$ in Section~\ref{sec:discussion}, where we also present our conclusions.
Some useful formulas and additional discussions are collected in the Appendix: 
Appendix~\ref{app:SO5approximatecase} contains a generalization of our derivation to theories where the strong dynamics contains a small breaking of the $SO(5)$ 
symmetry; the expressions of the spectral functions computed in the effective theory are reported in Appendix~\ref{app:spectralfunctions}; finally, 
in Appendix~\ref{app:example} we  illustrate a simple  model where the contribution to $\hat S$ from the lightest spin-1 resonances is not definite positive.

\section{Dispersion relation for $\eps_3$}
\label{sec:dispersionrelation} 

We start by deriving the dispersion relation for the $\eps_3$ parameter in the context of $SO(5)/SO(4)$ composite Higgs theories.
Our analysis will be  similar to that of Ref.~\cite{Orgogozo:2012ct}, although it differs in the way in which short- and long-distance contributions
from new physics are parametrized. 
In this respect our approach is closer to the original work of Peskin and Takeuchi~\cite{Peskin:1991sw}, where the $S$ parameter is defined to include
only short-distance effects from the new dynamics.

\subsection{Short- and long-distance contributions to $\eps_3$}

It is well known that universal corrections to the electroweak precision observables at the $Z$-pole can be described by three $\eps$ 
parameters~\cite{Altarelli:1990zd}. In this paper we are mainly interested in the $\eps_3$ parameter, 
which can be expressed as~\cite{Barbieri:1991qp}
\begin{equation}
\label{eq:eps3def}
\eps_3 = e_3 + c_W^2 e_4 - c_W^2 e_5 + (\text{non-oblique corrections})
\end{equation}
in terms of the vector-boson self energies
\begin{equation} \label{eq:selfenergies}
e_{3} =  \frac{c_W}{s_W}F_{3B}(m_{Z}^{2}),  \qquad
e_{4} =  F_{\gamma\gamma}(0)-F_{\gamma\gamma}(m_{Z}^{2}),  \qquad
e_{5} =  m_{Z}^{2}F_{ZZ}^{\prime}(m_{Z}^{2})\, .
\end{equation}
Here $s_W$ ($c_W$) denotes the sine (cosine) of the Weinberg angle and we have followed the standard convention decomposing the 
self energies (for canonically normalized gauge fields) as
\begin{equation}\label{eq:Pi}
\Pi_{ij}^{\mu\nu}(q) =  -i\eta^{\mu\nu} \left( A_{ij}(0) + q^2 F_{ij}(q^2) \right) + q^\mu q^\nu \text{ terms}\, .
\end{equation}
We consider scenarios in which the new physics modifies only the self energies, i.e. its effects are oblique. The  form of the non-oblique
vertex and box corrections in Eq.~(\ref{eq:eps3def}) is thus irrelevant to our analysis, since these cancel out when considering the new physics 
correction  $\Delta\eps_3 \equiv \eps_3 - \eps_3^{SM}$.
It is useful to distinguish between a short- and a long-distance contribution to $\Delta\eps_3$.
Heavy states with mass $m_* \gg m_Z$ affect only the short-distance part.
This latter can be expressed as the contribution of local operators, and is generated also by loops of light (i.e. Standard Model (SM)) particles. 
We define it to be
\begin{equation} \label{eq:deps3SD}
\Delta \eps_3|_{SD} = \Delta \bar e_3 + c_W^2 \Delta \bar e_4 - c_W^2 \Delta \bar e_5\, ,
\end{equation}
where $\Delta \bar e_i \equiv \bar e_i - \bar e_i^{SM}$ and 
\begin{equation} \label{eq:selfenergiesbar}
\bar e_{3} =  \frac{c_W}{s_W} \left( F_{3B}(0) + m_Z^2 F'_{3B}(0) \right),  \qquad
\bar e_{4} =  - m_Z^2 F^{\prime}_{\gamma\gamma}(0),  \qquad
\bar e_{5} =  m_{Z}^{2}F_{ZZ}^{\prime}(0)\, .
\end{equation}
It is convenient to express $\Delta\eps_3|_{SD}$ in terms of the  parameters $\hat S$, $W$, $Y$ and $X$ defined in Ref.~\cite{Barbieri:2004qk}:
\begin{equation} \label{eq:deps3SDSWYX}
\Delta\eps_3|_{SD} = \hat S - W - Y + \frac{X}{s_W c_W}\, ,
\end{equation}
where
\begin{equation} \label{eq:defS}
\begin{split}
\hat S & =  \frac{c_W}{s_W} ( F_{3B}(0) - F^{SM}_{3B}(0) )\\
X & = m_W^2 ( F^\prime_{3B}(0)-F^{\prime\, SM}_{3B}(0))  \, ,
\end{split}
\qquad
\begin{split}
W & = m_W^2 ( F^\prime_{WW}(0) - F^{\prime\, SM}_{WW}(0)) \\
Y & = m_W^2 ( F^\prime_{BB}(0) - F^{\prime\, SM}_{BB}(0) ) \, .
\end{split}
\end{equation}
The $S$ parameter originally introduced by Peskin and Takeuchi in Ref.~\cite{Peskin:1991sw} is related to $\hat S$ by $\hat S = (\alpha_{em}/4 s_W^2) S$.

\begin{figure}
\begin{center}
\includegraphics[width=0.30\textwidth]{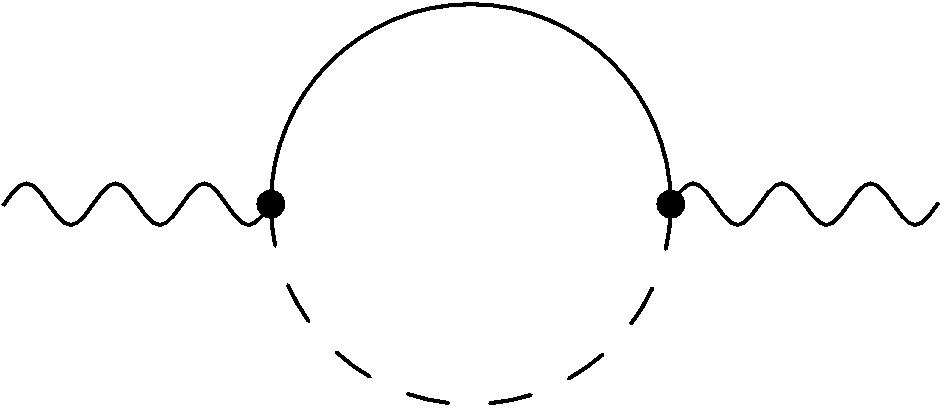}
\hspace{0.8cm}
\includegraphics[width=0.30\textwidth]{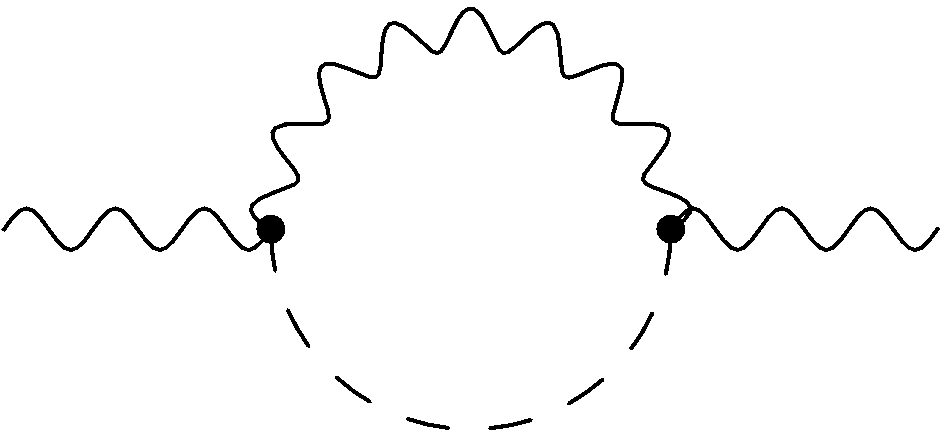}
\end{center}
\caption{\small
One-loop diagrams relative to the Higgs contribution to $\Delta\eps_3$. Wavy, continuous and dashed lines denote respectively gauge fields ($W^\pm$ and $Z$),
Nambu-Goldstone bosons of $SO(4)/SO(3)$ ($\pi^{1,2,3}$) and the Higgs boson.}
\label{fig:Higgscontribution}
\end{figure}

The long-distance correction to $\eps_3$ arises from loops of light particles only, as a consequence of their non-standard couplings.
We define
\begin{equation} \label{eq:deps3LD}
\Delta \eps_3|_{LD} = \left[ \Delta e_3 - \Delta \bar e_3 + c_W^2 (\Delta e_4 - \Delta \bar e_4) - c_W^2 ( \Delta e_5 - \Delta \bar e_5) \right]_\text{light particles}\, ,
\end{equation}
where $\Delta e_i \equiv e_i - e_i^{SM}$ and the expression in square brackets is computed by including only the contribution of light particles.
In the scenario under consideration the dominant long-distance contribution arises from the composite Higgs,
as a consequence of its modified couplings to vector bosons. At 1-loop it is given by the diagrams in Fig.~\ref{fig:Higgscontribution}. 
%
%
Working in the Landau gauge for
the elementary gauge fields ($\partial^\mu W^i_\mu = 0 = \partial^\mu B_\mu$), we find~\footnote{The same formula holds in a generic theory with Higgs coupling
to vector bosons  $c_V$ provided one replaces the factor $\sin^2\!\theta$ with $(1-c_V^2)$.}
\begin{equation} \label{eq:longdistance}
\begin{split}
\Delta\epsilon_3|_{LD} = \frac{g^2}{96\pi^2}\sin^2\!\theta \Bigg[ & f_3(x_h) -\frac{x_h}{2(1-x_h)^5}\left(x_h^4-5x_h^3+19x_h^2-9x_h+36\right)\log x_h \\
& - \frac{5x_h^4+7x_h^3+21 x_h^2+151 x_h+68}{12 (1-x_h)^4} \Bigg] \, ,
\end{split}
\end{equation}
where $x_h = m_h^2/m_Z^2$ and  the function $f_3$ is given by~\cite{Contino:2015mha}
\begin{equation}
\begin{split}
f_{3}(x) =& -x^{2}+3x-\frac{31}{6}+\frac{1}{4}\left(2x^{3}-9x^{2}+18x-12\right)\log x \\
                           & -\frac{\left(2x^{3}-13x^{2}+32x-36\right)x}{2\sqrt{(4-x)x}}\, \arctan\left(\sqrt{\frac{4}{x}-1}\right)\, .
\end{split}
\end{equation}
Additional long-distance effects arise from the top quark and are further suppressed by at least a factor $\zeta_t^2$,
where $\zeta_t$ is the degree of compositeness of the top quark. They will be neglected in the following.

From Eqs.~(\ref{eq:deps3SD}),~(\ref{eq:deps3SDSWYX}) and~(\ref{eq:deps3LD})  we find
\begin{equation}
\eps_3 = \eps_3^{SM} + \Delta \eps_3|_{LD} + \hat S - W - Y + \frac{X}{s_W c_W} + \dots \, .
\end{equation}
Together with Eq.~(\ref{eq:deps3LD}), this is our master formula for the calculation of $\eps_3$.~\footnote{An analogous
formula was given in Eq.~(6c) of Ref.~\cite{Barbieri:2004qk}, where however the long-distance term $\Delta \eps_3|_{LD}$ is omitted.}
It is accurate up to corrections (denoted by the dots)  of relative order $(m_Z^2/m_*^2)$, 
which are not captured by our definition of short- and long-distance contributions in Eqs.~(\ref{eq:deps3SD}) and~(\ref{eq:deps3LD}).
We will assume the mass scale of the new resonances to be much higher than the electroweak scale, $m_* \gg m_{Z}, m_h$, and neglect these corrections.

As a consequence of the gap between $m_*$ and $m_Z$, the contribution of the new heavy states to $\eps_3$ is local and encoded by the $\hat S, W, Y, X$ parameters.
Loops of light SM particles, in particular the Higgs boson, lead to an additional new physics correction through their modified couplings which is 
of both short- and long-distance types. In the composite Higgs theories under examination the shifts to the Higgs couplings are of order $(v/f)^2$, where $f$
is the Higgs decay constant. Since $f$ is related to $m_*$ through the coupling strength of the resonances, $m_* \sim g_* f$,
one could in principle get large modifications to the Higgs couplings for $f \sim v$ while still having a mass gap provided $g_* \gg g$.
In fact, current experimental data on Higgs production at the LHC disfavor large shifts and constrain 
$(v/f)^2 \lesssim 0.1$ at 95\% C.L.~\cite{Aad:2015pla} (see also Refs.~\cite{Azatov:2012qz,Falkowski:2013dza,Bellazzini:2014yua} for previous theoretical fits).
In the limit of a large compositeness scale, $f \gg v$, all the new physics contributions to low-energy observables
can be conveniently computed by matching the UV theory to an effective Lagrangian built with SM fields (including the Higgs doublet) at the scale $m_*$.
The leading contribution of light fields to $\Delta\eps_3$ then arises from 1-loop diagrams with one insertion of a dimension-6 operator.
The divergent part of these diagrams is associated with the RG running of the operators' coefficients, while the finite part is interpreted as a
long-distance threshold correction at the  scale $m_Z$.
This shows that the contributions from heavy modes and light modes are not individually RG invariant, as only their sum is independent of the 
renormalization scale at the one-loop level. Clearly, no issue with the RG invariance arises if one works at the tree level, and in that case it makes
perfect sense to define the $\hat S, W, Y$ and $X$ parameters to include only the contribution of heavy particles. When 1-loop corrections
are considered, however, any RG-invariant definition of the short-distance contribution  must include at least the divergent correction
from loops of light fields. According to our definition of Eq.~(\ref{eq:deps3SD}), $\hat S, W, Y$ and $X$ include such divergent part as well as a finite one.

\subsection{Dispersion relations for the short-distance contributions}
\label{subsec:dispersionforshortdistance}

We are now ready to  derive the dispersion relations  for $\hat S$, $W$ and $Y$ in terms of the spectral functions of the strongly-interacting dynamics.
We start by considering $\hat S$.

\begin{figure}
\begin{center}
\hspace{0.4cm}
\begin{minipage}[c]{0.25\textwidth}
\includegraphics[width=\textwidth]{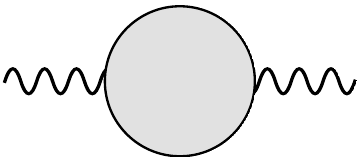}
\end{minipage}
\begin{minipage}[c]{0.085\textwidth}
\hspace{0.3cm} \vspace{-0.1cm} \large $+$ 
\end{minipage}
\begin{minipage}[c]{0.25\textwidth}
\vspace{-0.75cm}
\includegraphics[width=\textwidth]{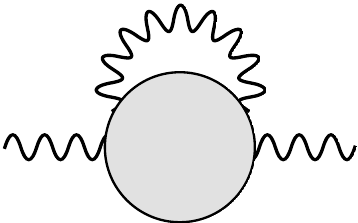}
\end{minipage}
\begin{minipage}[c]{0.2\textwidth}
\hspace{0.4cm} \vspace{-0.1cm} \large $+ \ \ \cdots$ 
\end{minipage}
\end{center}
\caption{\small
Contribution of the strong dynamics to the vector boson self energies expanded in powers of the weak gauge couplings. 
The gray blob in the first diagram corresponds to the correlator of two conserved currents of the strong dynamics.}
\label{fig:orders}
\end{figure}

The strong dynamics is assumed to have a global $SO(5)$ invariance 
spontaneously broken to $SO(4)\sim SU(2)_L \times SU(2)_R$.  The  elementary $W_\mu$ and $B_\mu$ fields 
gauge an $SU(2)_L \times U(1)_Y$ subgroup contained into an $SO(4)'$ misaligned by an angle $\theta$ with respect to the 
unbroken $SO(4)$ (see Refs.~\cite{Contino:2011np,Contino:2015mha} for details). 
They couple to the following linear combinations of $SO(5)$ currents~\footnote{We assume that the one in Eq.~(\ref{eq:Lint}) is the only interaction
between elementary gauge fields and the strong sector, i.e. that the gauge fields couple linearly to the strong dynamics through its conserved currents. If the
UV degrees of freedom of the strong dynamics include elementary scalar fields, then  an interaction  quadratic in the gauge fields is also present, as dictated by gauge
invariance.}
\begin{align} \label{eq:Lint}
{\cal L}_{int}  & = W^{a\, \mu} J_\mu^{a\, [W]} + B^\mu J_\mu^{[B]} \\[0.3cm]
\begin{split}
J_\mu^{a\, [W]} & = \Tr\!\left[  T^{aL}(0) T^A(\theta) \right] J^A_\mu \\
J_\mu^{[B]} & = \Tr\!\left[  T^{3R}(0) T^A(\theta) \right] J^A_\mu\, ,
\end{split}
\end{align}
where $T^A(\theta)$ are the $SO(5)$ generators, while $T^a(0)$ are the generators of the gauged $SO(4)'$.
Using the expressions for the generators given in Appendix A of Ref.~\cite{Contino:2011np} (see especially Eq.~(88) therein), we find
\begin{equation} \label{eq:currents}
\begin{split}
J_\mu^{3\, [W]} & = \left(  \frac{1+\cos\theta}{2} \right) J_\mu^{3L} + \left(  \frac{1-\cos\theta}{2} \right) J_\mu^{3R} + \frac{\sin\theta}{\sqrt{2}} J^{\hat 3}_\mu \\[0.2cm]
J_\mu^{[B]} & = \left(  \frac{1-\cos\theta}{2} \right) J_\mu^{3L} + \left(  \frac{1+\cos\theta}{2} \right) J_\mu^{3R} - \frac{\sin\theta}{\sqrt{2}} J^{\hat 3}_\mu \, ,
\end{split}
\end{equation}
where $J^{a_L}_\mu, J^{a_R}_\mu$ are the $SO(4)\sim SU(2)_L \times SU(2)_R$  currents ($a_L, a_R =1,2,3$) and $J^{\hat \imath}_\mu$  the $SO(5)/SO(4)$  ones 
($\hat \imath = 1,2,3,4$).
We assume that these currents are conserved in the limit in which the strong dynamics is taken in isolation, i.e. when the couplings to the elementary fields 
are switched off. This is for example the case of holographic composite Higgs models~\cite{holoCH}.
The generalization to the case in which the strong dynamics itself  contains a small source of explicit $SO(5)$ breaking is discussed in 
Appendix~\ref{app:SO5approximatecase}.
By working  at second order in the interactions (\ref{eq:Lint}) (i.e. at second order in the weak couplings), the vector-boson self energies 
in Eq.~(\ref{eq:defS}) can be  expressed in terms of two-point current correlators. 
The corresponding contribution to $\hat S$ and to the other oblique parameters $W, Y, X$ is gauge invariant~(see the detailed discussion in Ref.~\cite{Peskin:1991sw}).
The $\hat S$ parameter, in particular, gets a naive contribution of $O(m_Z^2/m_*^2)$ from the exchange of the heavy resonances of the strong dynamics,
while loops of Nambu-Goldstone (NG) bosons are responsible for the IR running of order $m_Z^2/(16\pi^2 f^2) \log(m_*/m_h)$.
Corrections from higher-order terms in the weak coupling expansion cannot be expressed as two-point current correlators and are not gauge invariant  in general.
A graphical representation of the various terms in the expansion is given in Fig.~\ref{fig:orders},
where a typical $O(g^4)$ contribution is exemplified by the second diagram.
%
%
A naive estimate shows that corrections at quartic order in the weak couplings from the exchange of heavy resonances are of order $m_Z^2/(16\pi^2 f^2) (g^2/g_*^2)$.
They are subdominant if $g \ll g_*$, and we will neglect them in the following.
In the case of corrections involving loops of  light fields only, on the other hand, the additional $g^2$ suppression can be compensated by inverse powers
of the light masses.  The only such unsuppressed contribution to $\hat S$ comes from the diagram on the right in Fig.~\ref{fig:Higgscontribution}, featuring a Higgs boson 
and a $Z$ in the loop. It is gauge invariant~\footnote{See the discussion in Ref.~\cite{Orgogozo:2012ct}.} and gives a correction
\begin{equation} \label{eq:SZh}
\delta\hat S_{Zh} = \frac{g^2}{96\pi^2}\frac{\sin^2\!\theta}{(x_h-1)^2} \left(\frac{9x_h+1}{2(1-x_h)}\log x_h + 2x_h + 3\right)\, ,
\end{equation}
which we will retain in our calculation.
Notice that since this term is not of the form of a two-point current correlator of the strong dynamics in isolation, it
was not included by Peskin and Takeuchi in their estimate of $S$ in Ref.~\cite{Peskin:1991sw}.~\footnote{For Technicolor one must set $\sin\theta = 1$ 
in Eq.~(\ref{eq:SZh}).}

In the limit in which the strong sector is taken in isolation, i.e. for unbroken $SO(5)$ symmetry, 
the Fourier transform of the Green functions of two conserved currents can be decomposed as:
\begin{equation} \label{eq:2pointGF}
\begin{split}
\langle J^{a_L}_\mu(q)  J_\nu^{b_L}(-q) \rangle  = &\, -i \delta^{a_L b_L} (P_T)_{\mu\nu} \, \Pi_{LL}(q^2)  \\[0.15cm]
\langle J^{a_R}_\mu(q)  J_\nu^{b_R}(-q) \rangle  = & \, -i \delta^{a_R b_R} (P_T)_{\mu\nu} \, \Pi_{RR}(q^2)  \\[0.15cm]
\langle J^{\hat a}_\mu(q)  J_\nu^{\hat b}(-q) \rangle  = & \,  -i \delta^{\hat a \hat b}  (P_T)_{\mu\nu}\, \Pi_{BB}(q^2)\, ,
\end{split}
\end{equation}
where $(P_T)_{\mu\nu} \equiv ( \eta_{\mu\nu} - q_\mu q_\nu/q^2 )$.
Any other two-point current Green function vanishes by $SO(5)$ invariance.
By using its definition in Eq.~(\ref{eq:defS}), together with Eqs.~(\ref{eq:Lint}),~(\ref{eq:currents})~and~(\ref{eq:2pointGF}), 
the parameter $\hat S$ can be expressed in terms of the correlators  $\Pi_{ij}$ as:
\begin{equation} \label{eq:Spi3B}
\hat S = g^2 \left( \Pi^\prime_{3B}(0) - \Pi^{hSM\, \prime}_{3B}(0) \right) + \delta\hat S_{Zh}\, ,
\end{equation}
where 
\begin{equation}
\label{eq:rel3B}
\Pi_{3B}(q^2) \equiv \frac{1}{4} \sin^2\!\theta \left( \Pi_{LL}(q^2) + \Pi_{RR}(q^2) - 2 \Pi_{BB}(q^2) \right)  \, ,
\end{equation}
and  $\Pi^{hSM}_{3B}$ denotes the expression of $\Pi_{3B}$ obtained by replacing the strong dynamics with the Higgs sector of the SM.
Equation~(\ref{eq:Spi3B}) is still a preliminary expression, however. 
The correlators $\Pi_{ij}(q^2)$ are singular at $q^2 =0$ due to the presence of the four massless NG bosons (including the Higgs boson), since they are computed
by considering the strong dynamics in isolation. 
A similar IR divergence is also present in the SM  Higgs sector, but only originating from the three $SO(4)/SO(3)$  NG bosons.
Subtracting the SM contribution in Eq.~(\ref{eq:Spi3B}) thus only partly removes the IR divergence.~\footnote{The IR divergence is completely removed if
the strong dynamics contains a small breaking of the $SO(5)$ symmetry giving the Higgs boson a mass. It is shown in Appendix~\ref{app:SO5approximatecase} 
that, even in this case,  it is useful to rewrite Eq.~(\ref{eq:Spi3B}) as discussed below to explicitly extract the Higgs chiral logarithm.}
There is, however, a simple way solve this problem and write a general formula for $\hat S$ in terms of two-point current correlators of the strong dynamics
in isolation.~\footnote{A possible alternative strategy is to define the correlators $\Pi_{ij}$ by including the explicit breaking of $SO(5)$ due to the coupling of  
the strong dynamics to the elementary fermions, in particular to the top quark. The resulting formula, however,  is less convenient to compute $\hat S$ by means of
non-perturbative tools such as lattice field theory. We thank Slava Rychkov for drawing our attention on the importance of working with 
two-point current correlators defined in terms of the strong sector in isolation.} 
Let us add and subtract in Eq.~(\ref{eq:Spi3B}) the contribution from a linear $SO(5)/SO(4)$ model defined in terms of the four NG bosons plus
an additional scalar field~$\eta$ which unitarizes the scattering amplitudes in the UV (see Appendix G of Ref.~\cite{Contino:2011np} for a definition). 
This model coincides with the $SO(5)/SO(4)$ strong dynamics in the infrared and is renormalizable.
Thus, we have:
\begin{equation} \label{eq:Sfinal}
\hat S = g^2 \left( \Pi^\prime_{3B}(0) - \Pi^{LSO5\, \prime}_{3B}(0) \right) + \delta \hat S_{LSO5} + \delta\hat S_{Zh}\, ,
\end{equation}
where $\Pi^{LSO5}_{3B}$ denotes the expression of $\Pi_{3B}$ obtained by replacing the strong dynamics with the linear $SO(5)/SO(4)$ model and
\begin{equation}
\label{eq:deltaLSO5}
\delta \hat S_{LSO5} \equiv g^2 \left(   \Pi^{LSO5\, \prime}_{3B}(0) - \Pi^{hSM\, \prime}_{3B}(0) \right) = \frac{g^2}{96\pi^2} \sin^2\!\theta \, \log\frac{m_\eta}{m_h}
\end{equation}
is computed for a non-vanishing Higgs mass. The mass of the scalar $\eta$ is an arbitrary parameter which can be taken to be of the order of the mass
of the heavy resonances of the strong sector, $m_\eta \sim m_*$.
In this way the Higgs chiral logarithm is fully captured by $\delta \hat S_{LSO5}$, and the first term in parenthesis in Eq.~(\ref{eq:Sfinal}) can be
evaluated setting the Higgs mass to zero (the relative error that follows is of order $m_h^2/m_*^2$ and can be thus neglected).
The IR singularities exactly cancel out in the difference of correlators in parenthesis, since the linear model by construction coincides 
with the strong dynamics in the infrared.
Equation (\ref{eq:Sfinal}), together with Eq.~(\ref{eq:rel3B}), is a generalization to $SO(5)/SO(4)$ composite Higgs theories of the analogous result derived 
in Ref.~\cite{Peskin:1991sw} by Peskin and Takeuchi for Technicolor.

At this point we can  make use of the dispersive representation of the correlators $\Pi_{ij}$.
This is obtained by inserting a complete set of states in the $T$-product of the two currents and defining 
\begin{equation} \label{eq:defspectral}
\sum_n \delta^{(4)}(q-q_n) \langle 0|J_\mu^i(0)| n\rangle \langle n|J_\nu^j(0)| 0\rangle =
 \frac{\theta(q_0)}{(2\pi)^3} \left( -\eta_{\mu\nu} q^2 \rho_{ij}(q^2) +  q_\mu q_\nu \,\bar \rho_{ij}(q^2) \right)\, .
\end{equation}
The spectral functions $\rho_{ij}$ and $(\bar\rho_{ij} - \rho_{ij})$  encode, respectively, the contribution of spin-1 and spin-0 intermediate states;
they are real and positive definite.  Current conservation implies $\rho_{ij} = \bar \rho_{ij}$, while 
from analyticity and unitarity it follows that 
\begin{equation}
\rho_{ij}(s) = \frac{1}{\pi} \text{Im}\!\left[ \frac{\Pi_{ij}(s)}{s} \right]\, .
\end{equation}
The $(n+1)$-subtracted dispersive representation thus reads (for a given $q^2_0$)
\begin{equation} \label{nsubdisprel}
\Pi_{ij}(q^2) = P_{n}(q^2)  + q^2 \left( q^2-q^2_0\right)^n \int_0^\infty  ds \, \frac{1}{(s-q^2_0)^n} \, \frac{\rho_{ij}(s)}{s-q^2+i\eps} \, ,
\end{equation}
where $P_{n}(q^2)$ is a polynomial of degree $n$.~\footnote{One has $P_0(q^2) =  \Pi_{ij}(0)$ and 
\begin{equation}
P_{n}(q^2) = \Pi_{ij}(0) \left( 1- \frac{q^2}{q_0^2} \right)^n  + 
q^2 \sum_{k=0}^{n-1} \frac{(q^2-q^2_0)^k}{k!} \frac{d^k}{d(q^2)^k} \left(\frac{\Pi_{ij}(q^2)}{q^2}\right)\!\bigg|_{q^2 = q^2_0} 
\quad (n \geq 1)\, .
\end{equation}
Notice that $\Pi_{LL}(0)$ and $\Pi_{RR}(0)$ vanish if the strong dynamics is considered in isolation.
}
It holds provided $\Pi_{ij}(q^2) \sim (q^2)^{1+n-\eps}$ for $|q^2| \to \infty$, with $\eps > 0$.
In the full theory of strong dynamics, the asymptotic behavior of the linear combination 
\begin{equation} \label{eq:Pi1}
\Pi_{1} \equiv \Pi_{LL} + \Pi_{RR} -2\Pi_{BB}
\end{equation}
is controlled by the scaling dimension, $\Delta \geq 1$, of the first scalar operator
entering its OPE (see the discussion in Ref.~\cite{Orgogozo:2012ct}): $\Pi_1(s) \sim s^{1-\Delta/2}$. One can thus write a dispersion representation for~$\Pi_1$ with 
just one subtraction (setting $n=0$ in Eq.~(\ref{nsubdisprel})), which in turn implies an unsubtracted  dispersive representation for $\hat S$.
Using the explicit expression of $\Pi^{LSO5\,\prime}_{3B}(0)$ we obtain:
\begin{equation} \label{eq:dispersionS}
\begin{split}
\hat S = \frac{g^2}{4}  \sin^2\!\theta \int_0^\infty \! \frac{ds}{s} \, \Bigg\{ 
 & \left(  \rho_{LL}(s) +  \rho_{RR}(s) - 2 \rho_{BB}(s) \right)   \\
 & - \frac{1}{48\pi^2} \left[  1 - \left(  1 - \frac{m_\eta^2}{s} \right)^3 \theta(s - m_\eta^2) \right]
\Bigg\} + \delta \hat S_{LSO5} + \delta\hat S_{Zh} \, .
\end{split}
\end{equation}
%
This result generalizes the dispersion formula derived by Peskin and Takeuchi in Ref.~\cite{Peskin:1991sw} for Technicolor to the case of $SO(5)/SO(4)$
composite Higgs theories. The dispersive integral accounts for the contribution from heavy states (of $O(m_Z^2/m_*^2)$), while the chiral logarithm
due to Higgs compositeness is encoded by $\delta \hat S_{LSO5}$. The dependence on $m_\eta$ cancels out when summing this latter term with the
dispersive integral.

Let us now turn to $W$, $Y$ and $X$.
In our class of theories the contribution of heavy particles to $X$ is of $O(m_Z^4/m_*^4)$ and will be neglected (it is of the same order as the uncertainty
due to our definition of short- and long-distance parts in $\Delta\eps_3$).
The contribution of heavy particles to $W$ and $Y$ is instead of $O[(m_Z^2/m_*^2)(g^2/g_*^2)]$ and will be retained.
Finally, the contribution to $W$, $Y$ and $X$ from the diagrams of Fig.~\ref{fig:Higgscontribution} involving light particles only 
is not suppressed and must be fully included.
For $X$ we  find
\begin{equation}
X = -\frac{g^2}{64\pi^2} s_W c_W \sin^2\!\theta \left[  \frac{3 x_h^2 +4 x_h}{(x_h -1)^5} \log x_h - \frac{x_h^3 + x_h^2 + 73 x_h +9}{12 (x_h-1)^4}  \right]  + \dots
\end{equation}
where the dots indicate $O(m_Z^4/m_*^4)$ terms generated by the exchange of heavy particles.
In the case of $W$ and $Y$, it is straightforward to derive a dispersion relation by following a procedure analogous to that discussed for $\hat S$.~\footnote{The dispersive
representation of $\Pi_{LL}$ and $\Pi_{RR}$ in this case requires two subtractions ($n=1$ in Eq.~(\ref{nsubdisprel})), since $\Pi_{LL}(q^2)\sim \Pi_{RR}(q^2) \sim q^2$ 
for $|q^2| \to \infty$.}
By neglecting terms of order $O(m_W^4/m_*^4)$, we obtain~\footnote{The $O(m_W^4/m_*^4)$ neglected terms give a contribution to $W$ which can be written
as follows:
\label{fot:neglectedterms}
\begin{equation}
\begin{split}
\delta W = m_W^2 g^2 \Bigg\{ 
       &  - \frac{\sin^2\theta}{4} \int^\infty_0 \! \frac{ds}{s^2}\, \Bigg[ \left( \rho_{LL}(s) + \rho_{RR}(s)  - 2 \rho_{BB}(s) \right)    \\
       & \phantom{- \frac{\sin^2\theta}{4} \int^\infty_0 \! \frac{ds}{s^2}\, \Bigg[ \, } - \frac{1}{48\pi^2} 
           \left(  1 - \left(  1 - \frac{m_\eta^2}{s} \right)^3 \theta(s - m_\eta^2) \right) \Bigg] \\[0.1cm]
       &  - \sin^2\frac{\theta}{2} \int^\infty_0 \! \frac{ds}{s^2}\, \left( \rho_{LL}(s) - \rho_{RR}(s) \right) \Bigg\} \, .
\end{split}
\end{equation}
The additional contribution to $Y$ has the same form provided one exchanges $LL \leftrightarrow RR$ and $g \leftrightarrow g'$.
}:
\begin{align}
\label{eq:dispersionW}
W = m_W^2 g^2  \int^\infty_0 \! \frac{ds}{s^2}\, \left( \rho_{LL}(s) - \frac{1}{96\pi^2} \right) - \frac{g^2}{96\pi^2} \frac{c_W^2}{8 x_h} \sin^2\!\theta  + \delta W_{Zh} \\[0.3cm]
\label{eq:dispersionY}
Y = m_W^2 g'^2 \int^\infty_0 \! \frac{ds}{s^2}\, \left( \rho_{RR}(s) - \frac{1}{96\pi^2} \right) - \frac{g'^2}{96\pi^2} \frac{c_W^2}{8 x_h} \sin^2\!\theta  + \delta Y_{Zh}\, .
\end{align}
The first term in each equation  encodes the contribution from the heavy resonances and is of $O[(m_Z^2/m_*^2)(g^2/g_*^2)]$.
In particular, the  integral in Eq.~(\ref{eq:dispersionW}) equals $(\Pi_{LL}^{\prime\prime}(0) - \Pi_{LL}^{LSO5\, \prime\prime}(0) )$, 
while that in Eq.~(\ref{eq:dispersionY}) equals  $(\Pi_{RR}^{\prime\prime}(0) - \Pi_{RR}^{LSO5\, \prime\prime}(0))$.
The second terms come from the difference between the $SO(5)/SO(4)$ linear model and the SM (they are the analogous to Eq.~(\ref{eq:deltaLSO5})), while
$\delta W_{Zh}$ and $\delta Y_{Zh}$ are the contributions from the $Zh$ loop in Fig.~\ref{fig:Higgscontribution}:
\begin{equation}
\delta W_{Zh} = \frac{g^2}{g'^2} \, \delta Y_{Zh}  = 
 \frac{g^2}{64\pi^2} c^2_W \sin^2\!\theta \left[  \frac{3 x_h^2 +4 x_h}{(x_h -1)^5} \log x_h - \frac{5 x_h^3 + 67 x_h^2 + 13 x_h -1}{12 x_h (x_h-1)^4}  \right] \, .
\end{equation}

By putting together the expressions of $\hat S$, $W$, $Y$, $X$, and of the long-distance part Eq.~(\ref{eq:longdistance}), 
we obtain a dispersive formula for $\Delta\eps_3$:
\begin{equation}
\label{eq:dispersioneps3}
\begin{split}
\Delta\eps_3 = & \, \frac{g^2}{96\pi^2} \sin^2\!\theta \left(f_3(x_h) + \frac{\log x_h}{2}  - \frac{5}{12}  + \log\frac{m_\eta}{m_h}  \right)  \\
 & + \frac{g^2}{4}  \sin^2\!\theta \int_0^\infty \! \frac{ds}{s} \, \Bigg\{ 
     \rho_{LL}(s) +  \rho_{RR}(s) - 2 \rho_{BB}(s)   \\
 & \phantom{+ \frac{g^2}{4}  \sin^2\!\theta \int_0^\infty \! \frac{ds}{s} \, \Bigg\{ \, } 
    - \frac{1}{48\pi^2} \left[  1 - \left(  1 - \frac{m_\eta^2}{s} \right)^3 \theta(s - m_\eta^2) \right] \Bigg\}  \\
 & + m_W^2  \int^\infty_0 \! \frac{ds}{s^2}\, \left( g^2  \rho_{LL}(s) + g'^2  \rho_{RR}(s) -  \frac{g^2+g'^2}{96\pi^2}  \right) \, .
\end{split}
\end{equation}
The second and third terms encode the contribution from the heavy resonances and are, respectively, of $O(m_Z^2/m_*^2)$ and $O[(m_Z^2/m_*^2)(g^2/g_*^2)]$.
When modeling the spectral functions  --as we will do in the next section-- in terms of the lowest-lying resonances of the strong dynamics, these contributions arise 
from the tree-level exchange of  massive spin-1 states.
We neglected terms of $O(m_Z^4/m_*^4)$ (arising in particular from our definition of short- and long-distance contributions) and of $O[(m_W^2/16\pi^2 f^2) (g^2/g_*^2)]$
(arising from the expansion in powers of the weak  couplings required to obtain a formula in terms of current correlators).

Equation~(\ref{eq:dispersioneps3}) should be compared to the analogous result previously derived by Rychkov and Orgogozo in Ref.~\cite{Orgogozo:2012ct}.
The expression given there also relies on an expansion in $g^2$ and does not include the heavy-particle contribution to $W$ and $Y$ (the last term of 
our Eq.~(\ref{eq:dispersioneps3})). 
Rychkov and Orgogozo also define the dispersive integral to comprise the contribution of the heavy states only, but do not perform any subtraction to remove
the NG boson contribution. Rather,  the integration over light modes is done explicitly and in an approximate way.
Their procedure implies a relative uncertainty of  order $m_h/m_*$, which follows in particular from neglecting the Higgs mass
and the contribution of the heavy states in the evaluation of the low-energy part of the dispersive integral.
In our case the relative uncertainty implied by our definition of  short- and long-distance parts is smaller and of order $(m_Z/m_*)^2$.
Within their accuracy, the two results coincide.

\section{Dispersive relation in the effective theory}
\label{sec:disersiveEFT} 

The dispersive integrals in Eq.~(\ref{eq:dispersioneps3}), as well as those in Eqs.~(\ref{eq:dispersionS}),~(\ref{eq:dispersionW}) and~(\ref{eq:dispersionY}), 
are convergent and well defined if the spectral functions are computed in the full theory of the strong dynamics.
Here we want to provide an approximate calculation of $\Delta\eps_3$ which makes use of an effective description of the strong dynamics in terms
of its lowest-lying resonances and NG bosons. We focus in particular on the contribution of a spin-1 resonance ($\rho_L$) transforming as
a $(3,1)$ of the $SO(4) \sim SU(2)_L \times SU(2)_R$ global symmetry. We will thus compute the spectral functions in the effective theory and
integrate them to obtain $\hat S$, $W$ and $Y$, hence $\Delta\eps_3$, through their dispersion relations. In this case, the spectral
integrals are generically divergent  in the ultraviolet,  since the effective description is
approximately valid at low energy but not adequate for momenta larger than the cutoff scale. In other words, the dispersion relations
derived in the previous section need to be modified in order to be used in the effective theory. Let us see how.

By considering the gauge fields $A_\mu$ as external sources for the currents,  any two-point current correlator can be  expressed as the second derivative of an 
effective action $W[A]$ with respect to the  source:
\begin{equation} \label{eq:2pointW}
\langle J_\mu(x)  J_\nu(y) \rangle  = (-i)^2 \frac{\delta^2}{\delta A^\mu(x) \delta A^\nu(y)} W[A]\bigg|_{A=0}\, ,
\end{equation}
where
\begin{equation}
W[A] = \log \int\! d\varphi \, \exp\left(i S[\varphi] + i \int \! d^4x\,   J_\mu A^\mu\right)
\end{equation}
and $\varphi$ denotes the UV degrees of freedom of the strong dynamics.
In the absence of a description of the theory in terms of these fields, we can compute $W[A]$ approximately as the integral over the IR degrees of freedom $\varphi_{IR}$:
\begin{equation} \label{eq:WIR}
W[A] \simeq \log \int\! d\varphi_{IR} \, \exp\left(i S_{IR}[\varphi_{IR}, A]\right)\, .
\end{equation}
Notice however that the low-energy action $S_{IR}$ will not depend on the source only through its coupling to the low-energy conserved current $J_{\mu}^{IR}$,
but will contain non-minimal interactions. At  quadratic order in the source, we can write
\begin{equation} \label{eq:SIR}
S_{IR}[\varphi_{IR}, A] = S_{IR}[\varphi_{IR}] + \int\! d^4x\, \left( J_{\mu}^{IR} A^\mu + O_{\mu\nu} A^{\mu\nu} + \frac{c_0}{2}  A_\mu A^\mu 
 - \frac{c_1}{4} \, A_{\mu\nu} A^{\mu\nu} +\dots \right)
\end{equation}
where $c_{0}$ and $c_1$ are constants, $A_{\mu\nu}$ is the field strength constructed with the source and $O_{\mu\nu}$ is an operator
antisymmetric in its Lorentz indices.
The second term in the parentheses is a non-minimal interaction 
that is generated when flowing to the infrared. The last two terms in parentheses depend only on the source and generate contact contributions upon
differentiation; pure-source higher-derivative terms are denoted by the dots.
By using Eqs.~(\ref{eq:SIR}) and~(\ref{eq:WIR}) to compute (\ref{eq:2pointW} one finds
\begin{equation}
\langle J_\mu(x)  J_\nu(y) \rangle = \langle \tilde J_\mu(x)  \tilde J_\nu(y) \rangle + c_0 \, \eta_{\mu\nu}  \delta^{(4)}(x-y)
  + c_1 \left( \eta_{\mu\nu} \Box - \partial_\mu\partial_\nu \right) \delta^{(4)}(x-y)+\dots \, ,
\end{equation}
where $\tilde J_\mu \equiv J_\mu^{IR} - 2\, \partial^\rho O_{\rho\mu}$ is also a conserved current, and the dots stand for higher-derivative local terms.
The Green functions $\langle J_\mu  J_\nu \rangle$ can thus be computed in terms of the two-point functions of the effective currents $\tilde J_\mu$.
The coefficients $c_i$ are arbitrary in the effective theory and can be chosen to cancel the UV divergences arising in 
$\langle \tilde J_\mu  \tilde J_\nu \rangle$.~\footnote{The value of $c_0$ can be adjusted 
to ensure that the contributions to the two-point correlator from the tree-level exchange of, respectively, one NG boson
and one spin-1 resonance are transverse. A simple way to enforce the Ward identity is in fact demanding that the effective action $S_{IR}[\varphi_{IR}, A]$
be invariant under local $SO(5)$ transformations under which the source $A_\mu$  transforms as a gauge field. We thank Massimo Testa for a discussion on this point.
Notice also that adding the pure source terms in Eq.~(\ref{eq:SIR}) corresponds to a redefinition of the $T^*$ product of  two currents.}
Performing a Fourier transformation one has
\begin{equation}
\label{eq:PivsPitilde}
\Pi_{ij}(q^2) = \tilde \Pi_{ij}(q^2) + \Delta_{ij}(q^2)\, ,
\end{equation}
where $\tilde \Pi_{ij}$ is the two-point current correlator in the effective theory and $\Delta(q^2) = \sum_k (q^2)^k c_k$ denotes the local counterterms.

It is always possible to express $\tilde \Pi_{ij}(q^2)$ as an integral over a contour in the complex plane that
runs below and above its branch cut on the real axis (where the imaginary part of $\tilde \Pi_{ij}$ is discontinuous) 
and then describes a circle of radius $M^2$ counterclockwise.
We thus obtain
\begin{equation}
\label{eq:dispersionMfinite}
\Pi_{ij}(q^2) = \tilde\Pi_{ij}(0) + q^2 \int_0^{M^2} \!\! ds\; \frac{\tilde\rho_{ij}(s)}{s-q^2} + \frac{q^2}{2\pi i} \int_{C_{M^2}} \!\! dz\; \frac{\tilde\Pi_{ij}(z)}{z (z-q^2)} 
 + \Delta_{ij}(q^2)\, ,
\end{equation}
where $C_{M^2}$ denotes the part of the contour over the circle, and $\tilde \rho_{ij}(q^2) = (1/\pi) \text{Im}[ \tilde\Pi_{ij}(q^2)/q^2 ]$ is the spectral function of the currents 
$\tilde J_\mu$. Since the value of $M$ is arbitrary (as long as $q^2$ is inside the contour),  the dependence on $M^2$ cancels out  in Eq.~(\ref{eq:dispersionMfinite}).
If $\tilde \Pi_{ij}(q^2)/q^2\to 0$ for $|q^2|\to \infty$,  it is possible to take the limit $M^2\to \infty$ so that the integral on the
circle vanishes. In this case one obtains a dispersion relation for $\Pi_{ij}(q^2)$ in terms of $\tilde \rho_{ij}$ similar to the one valid in the full theory, except
for the appearance of the local term. In general, however, the correlator $\tilde\Pi_{ij}$ is not sufficiently well behaved at infinity, and $M$ must be kept finite.
If $\tilde\Pi(q^2) \sim (q^2)^{1+k}$ at large $q^2$, both the dispersive integral and the integral over the circle scale as $(M^2/m_*^2)^k$, where $m_*$ is
the mass of the resonances included in the low-energy theory.
Also,  $\tilde\Pi_{ij}$ generally requires a regularization to be defined and contains divergences which 
are removed by the counterterm $\Delta_{ij}$. The dispersive integral, on the other hand, is convergent since $\tilde\rho_{ij}$ is finite (after subdivergences
are removed).

A particularly convenient way to define $\tilde\Pi_{ij}(q^2)$ is through dimensional regularization. Upon  extending the theory to $D$ dimensions, indeed, its 
asymptotic $q^2$  behavior arising at the radiative level can be arbitrarily softened.
For example,  the 1-loop contribution to $\tilde\Pi_{ij}$ scales like $(q^2)^{1+n-\eps/2}$ at large $q^2$, where $n$ is some integer and $\eps \equiv 4-D$.
It is thus  possible to choose~$\eps$ sufficiently large and positive ($\eps > 2n$), such that the contribution to the integral on the circle from 1-loop effects 
 vanishes when taking the limit $M^2 \to \infty$.  
In doing so, the dispersive integral (now with its upper limit extended to infinity) becomes singular for $\eps\to 0$. 
The divergence is thus transferred from the integral over the circle to the dispersive integral, and the $1/\eps$ poles are still removed by the counterterm $\Delta_{ij}$.
The same argument goes through after including higher-loop contributions. The large-$q^2$ behavior of the tree-level part of $\tilde\Pi_{ij}$, on the other hand,
cannot be softened through dimensional continuation. If thus $\tilde\Pi_{ij}$ scales like $(q^2)^{1+n}$ at tree level, with $n > 0$, it is not possible to take the
$M^2 \to \infty$ limit in Eq.~(\ref{eq:dispersionMfinite}) (unless one performs $n$ additional subtractions). The case with $n=0$ is special, in that $M^2$ can be sent to
infinity but the integral over the circle tends to a constant and does not vanish.
Assuming that $\tilde\Pi_{ij}(q^2)$ grows no faster than $q^2$ in $D$ dimensions,
one can thus derive the following dispersion relation:
\begin{equation}
\label{eq:dispersionMinfinite}
\Pi_{ij}(q^2) = \tilde\Pi_{ij}(0) + q^2 \int_0^{\infty} \!\! ds\; \frac{\tilde\rho_{ij}(s)}{s-q^2} + \Delta_{ij}(q^2) + q^2 C_{ij}\, ,
\end{equation}
where
\begin{equation}
\label{eq:deltainfty}
C_{ij} \equiv \lim_{|q^2|\to \infty} \left[ \frac{\tilde\Pi_{ij}(q^2)}{q^2} \right]\, .
\end{equation}
This is the formula that we will use in the next section to compute $\hat S$, $W$ and $Y$.

We conclude by noticing that another approach is also possible to derive a dispersion relation in the effective theory.
One could use Eq.~(\ref{eq:PivsPitilde}) and approximate $\text{Im}[\Pi_{ij}(q^2)] \simeq \text{Im}[\tilde\Pi_{ij}(q^2)]$ for $q^2 \ll \Lambda^2$.
Substituting $\rho_{ij}(s) = \tilde\rho_{ij}(s) + O(s/\Lambda^2)$ in the dispersion relation of the full theory, one thus obtains
\begin{equation} \label{eq:approach2}
\Pi_{ij}(q^2) = \tilde\Pi_{ij}(0) + q^2 \int_0^{M^2} \!\! ds\; \frac{\tilde\rho_{ij}(s)}{s-q^2} + q^2 \int_{M^2}^\infty \!\! ds\; \frac{\rho_{ij}(s)}{s-q^2} 
                     + O\!\left(\frac{M^2}{\Lambda^2}\right)\, .
\end{equation}
The value of $M$ can be conveniently chosen to be much larger than the  mass of the resonances $m_*$, so as to fully include their
contribution to the dispersive integral, and much smaller than the cutoff scale $\Lambda$, as required for $\tilde \rho_{ij}$ to give a good approximation
of the full spectral function. With this choice,
the last two terms in Eq.~(\ref{eq:approach2}) encode the contribution from the cutoff dynamics.
Comparing with Eq.~(\ref{eq:dispersionMfinite}), it follows that 
\begin{equation}
q^2 \int_{M^2}^\infty \!\! ds\; \frac{\rho_{ij}(s)}{s-q^2}  = 
   \frac{q^2}{2\pi i} \int_{C_{M^2}} \!\! dz\; \frac{\tilde\Pi_{ij}(z)}{z (z-q^2)} + \Delta_{ij}(q^2) + O\!\left(\frac{M^2}{\Lambda^2}\right) \, .
\end{equation}
%

\section{One-loop computation of $\Delta\eps_3$}
\label{sec:calculation} 

Having discussed how the dispersion relations are modified in the effective theory, we now 
put them to work and perform an explicit calculation of $\Delta\eps_3$.
Our goal is thus computing the spectral functions $\tilde\rho_{ij}$ of the currents $\tilde J_\mu$ 
in the effective theory with NG bosons and a spin-1 resonance $\rho_L$.
The dynamics of the spin-1 resonance will be described by the effective Lagrangian of Ref.~\cite{Contino:2015mha} (see Eqs.~(2.6) and (2.16) therein), 
the notation of which we follow.
The $SU(2)_L$, $SU(2)_R$ and $SO(5)/SO(4)$ components of $\tilde J_\mu$ read, respectively:
\begin{align}
\label{eq:JtildeL}
\tilde J_\mu^{a_L} & = \frac{(1-a_\rho^2)}{2}  \left( \epsilon^{a_L bc}\partial_\mu \pi^b \pi^c + \partial_\mu \pi^{a_L} \pi^4 - \partial_\mu \pi^4 \pi^{a_L} \right)
  - \frac{m_\rho^2}{g_\rho} \rho_\mu^a -2 \alpha_2 g_\rho \partial^\alpha \rho_{\alpha\mu} + \dots   \\
\tilde J_\mu^{a_R} & = \frac{1}{2}  \left( \epsilon^{a_R bc}\partial_\mu \pi^b \pi^c + \partial_\mu \pi^{a_R} \pi^4 - \partial_\mu \pi^4 \pi^{a_R} \right) + \dots  \\
\tilde J^{\hat{a}}_\mu & = \frac{f}{\sqrt{2}} \partial_\mu \pi^{\hat a} - \frac{f}{\sqrt{2}} a_\rho^2 g_\rho 
 \left( \epsilon^{abc} \rho_\mu^b \pi^c + \delta^{\hat a4} \rho_\mu^b \pi^b - \rho_\mu^{\hat a} \pi^4 \right) + \dots
\end{align}
where $g_\rho$ is the resonance's coupling strength, $a_\rho \equiv m_\rho/(g_\rho f)$ and the ellipses denote terms with higher powers of the fields or terms that 
are not relevant for the present calculation.
The last term in Eq.~(\ref{eq:JtildeL}) proportional to~$\alpha_2$ originates from the non-minimal coupling to the external source induced by the 
operator $Q_2 = \Tr[ \rho_L^{\mu\nu} f^L_{\mu\nu}]$.~\footnote{Notice that a different basis was used in Ref.~\cite{Contino:2015mha} where
$Q_2 = \Tr[ \rho_L^{\mu\nu} E^L_{\mu\nu}]$. The definition adopted in this paper is more convenient for our discussion.}

\begin{figure}[t]
\begin{center}
\begin{minipage}[c]{0.14\textwidth}
\centering
$\tilde\rho_{LL}$:
\end{minipage}
\begin{minipage}[c]{0.85\textwidth}
\includegraphics[height=0.17\textwidth]{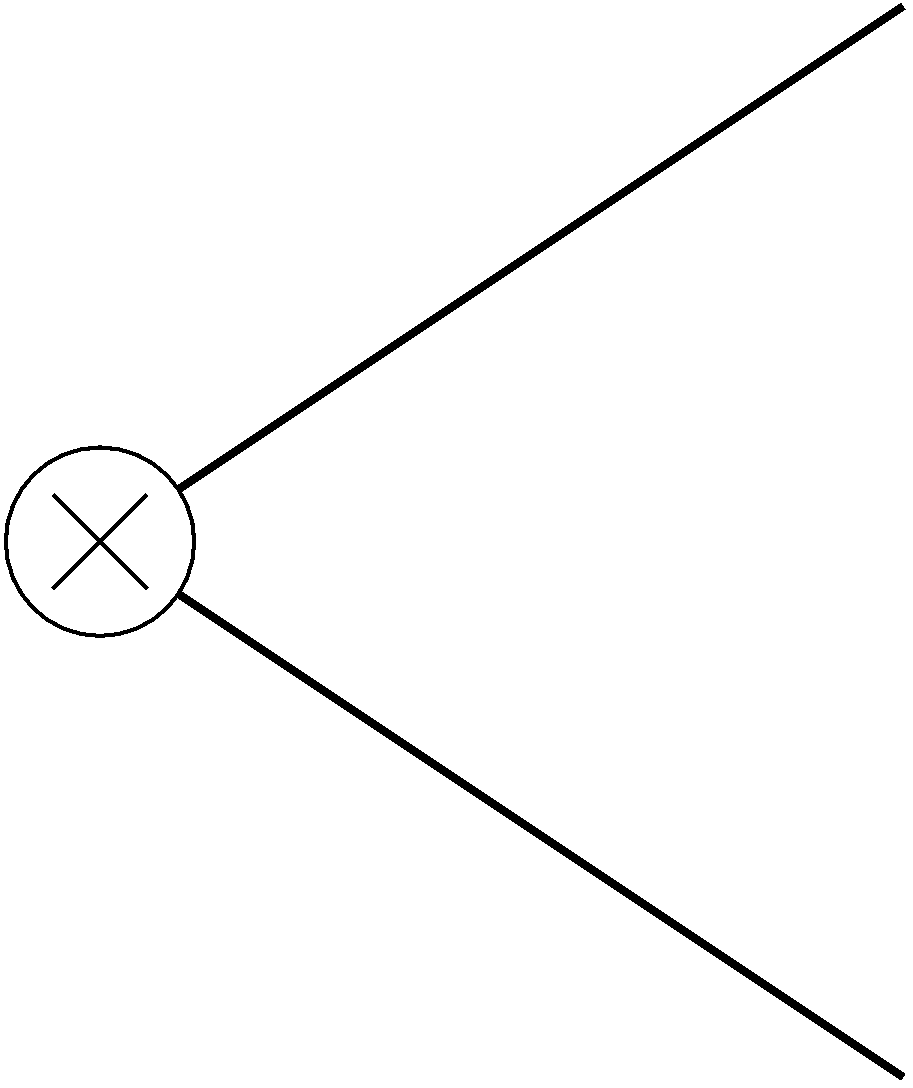}
\hspace{0.8cm}
\includegraphics[height=0.17\textwidth]{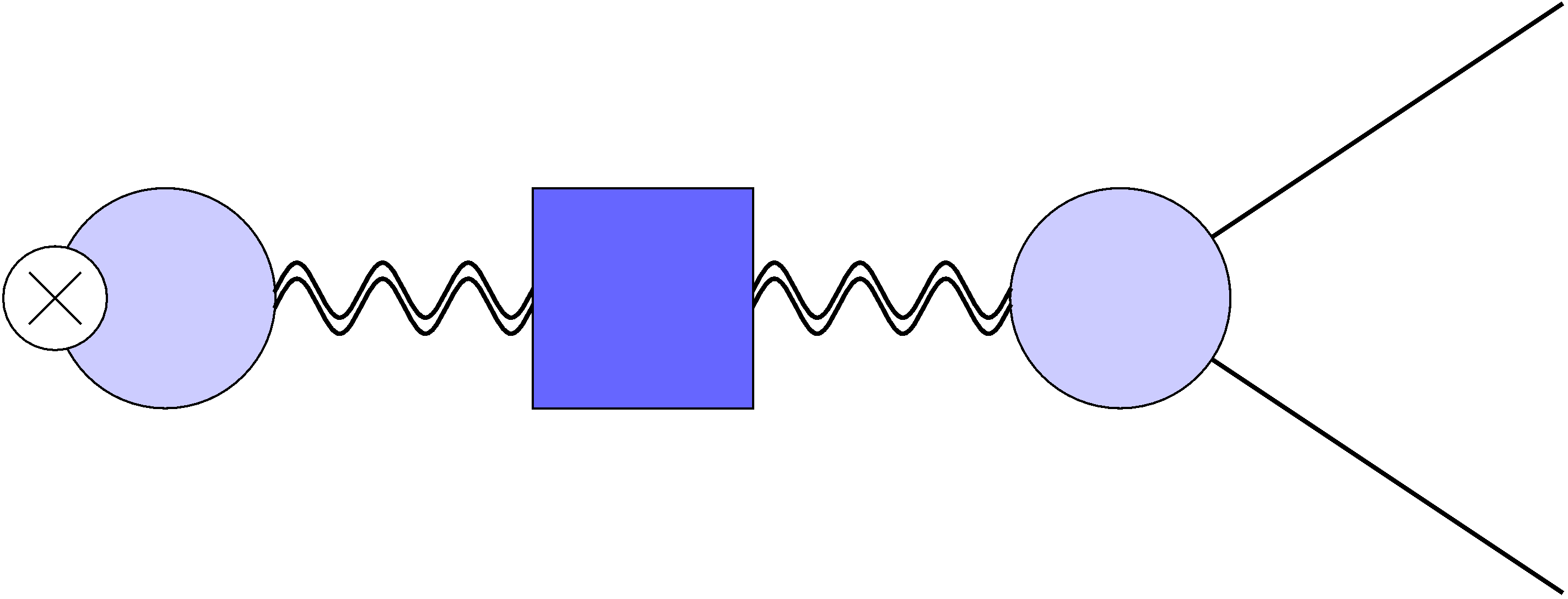}
\hspace{0.8cm}
\includegraphics[height=0.17\textwidth]{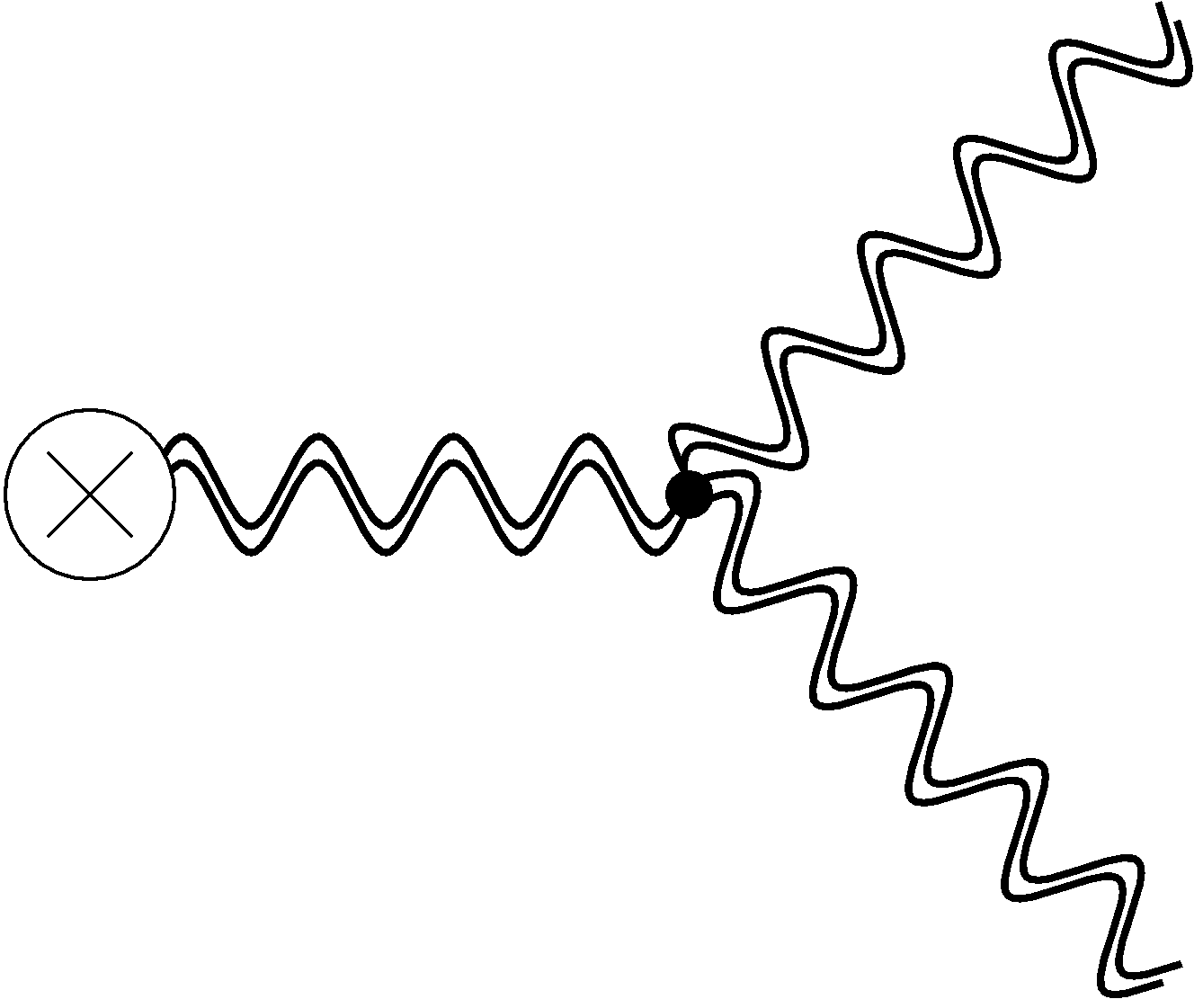}
\end{minipage}
\\[0.9cm]
\begin{minipage}[c]{0.14\textwidth}
\centering
$\tilde\rho_{RR}$:
\end{minipage}
\begin{minipage}[c]{0.85\textwidth}
\includegraphics[height=0.17\textwidth]{NewPlots/JU_contact.png}
\end{minipage}
\\[0.9cm]
\begin{minipage}[c]{0.14\textwidth}
\centering
$\tilde\rho_{BB}$:
\end{minipage}
\begin{minipage}[c]{0.85\textwidth}
\includegraphics[height=0.17\textwidth]{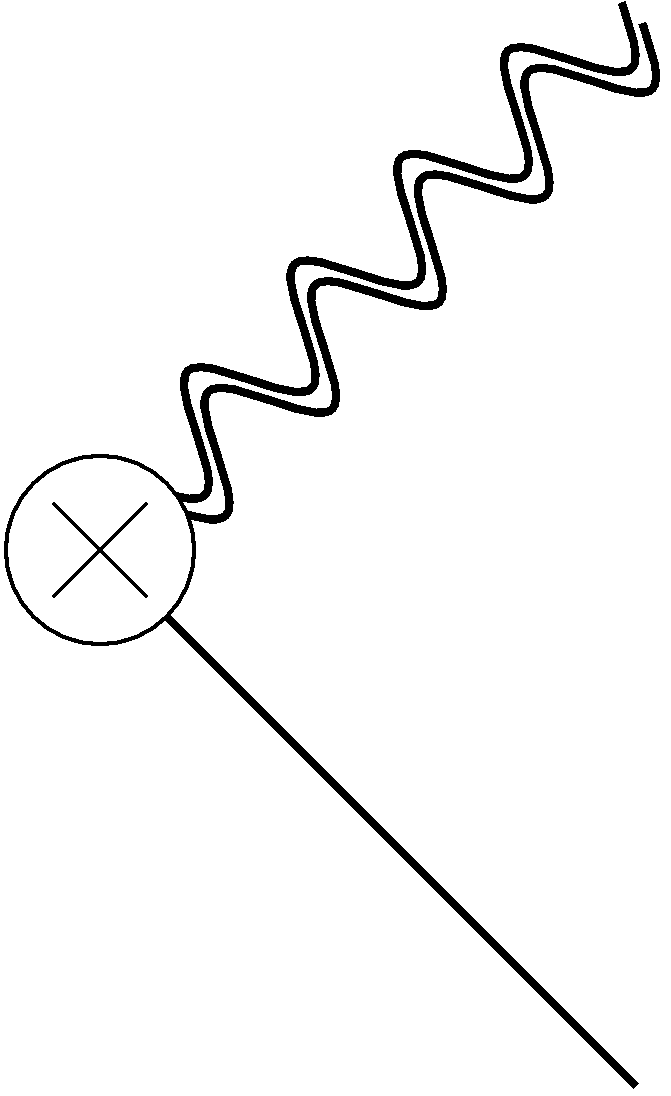}
\hspace{1.32cm}
\includegraphics[height=0.17\textwidth]{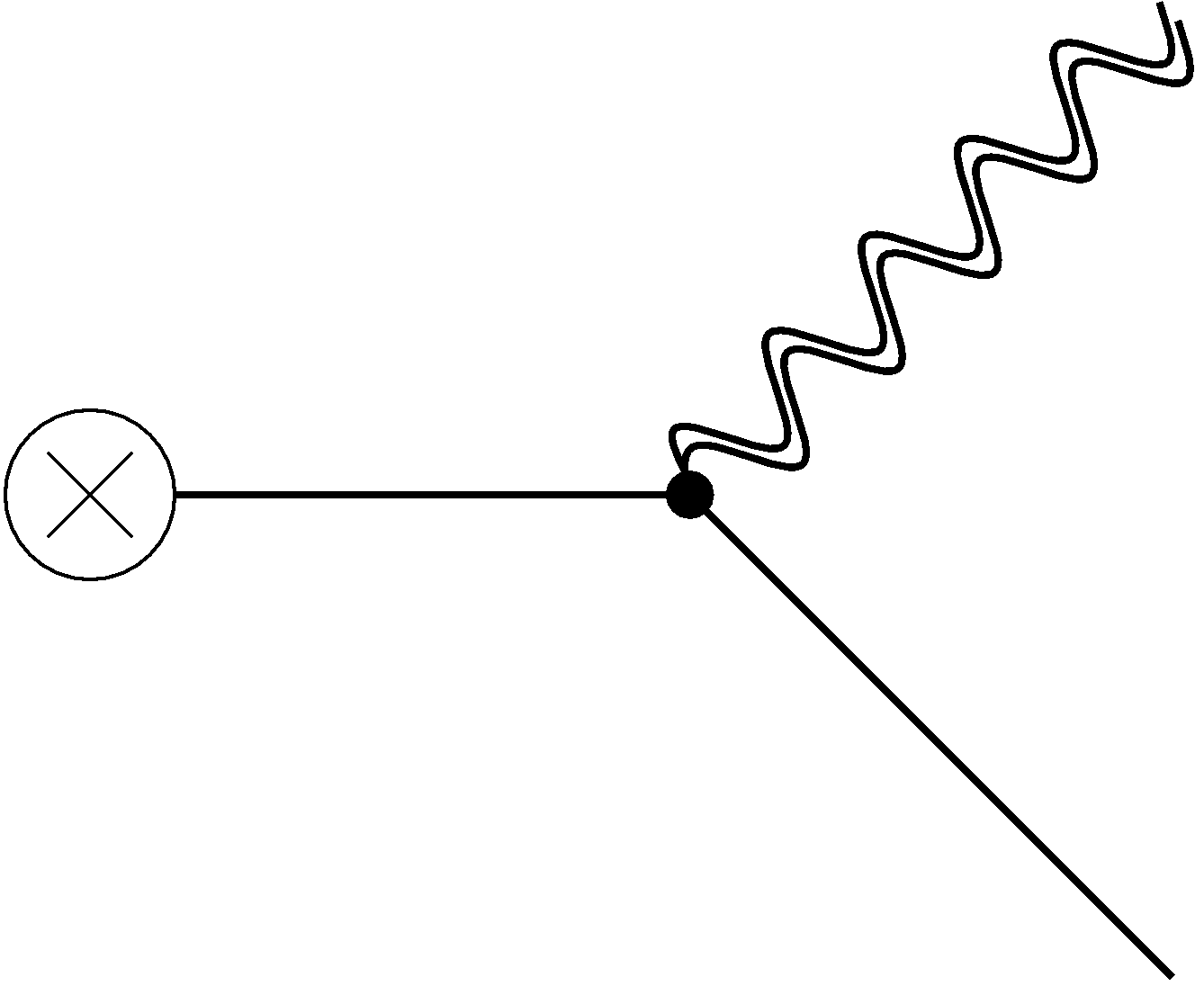}
\end{minipage}
\end{center}
\caption{\small Feynman diagrams contributing to the spectral functions in the effective theory at $O(g_\rho^0)$. Continuous lines denote NG bosons of $SO(5)/SO(4)$ 
($\pi^{\hat a}$), while double wavy lines denote a $\rho_L$. The cross stands for the insertion of a current, while the blue blobs and box indicate 
respectively the 1-loop corrected vertices and propagator.}
\label{fig:diagrams}
\end{figure}

To compute the spectral functions, we use the definition~(\ref{eq:defspectral}) in terms of a sum over intermediate states.
The resonance $\rho_L$ can decay to two NG bosons and is not an asymptotic state. The intermediate states to be considered are thus multi-NGB
states:~\footnote{The  exchange of one NG boson contributes only to the spectral function $\bar\rho_{BB}$ and is thus irrelevant to our calculation.}
$\pi\pi$, $3\pi$, $4\pi$, $\dots$. 
It is however possible to simplify the calculation by noticing the following. We want to derive an expression for the $\hat S$ parameter at order $g^0_\rho$,
by expanding for $g_\rho/4\pi$ small. Since the  contribution from the tree-level exchange of the $\rho_L$ is of order $1/g_\rho^2$,  our result will include
terms that appear at the 1-loop level in a diagrammatic calculation of $\hat S$. 
The role of tree- and loop-level effects in the dispersive computation, on the other hand, is subtler.
Consider for example the contribution to the $\pi\pi$ state coming from the exchange of a $\rho_L$, i.e. 
that of the second diagram in the first row of Fig.~\ref{fig:diagrams}.
%
%
The vertex  with the current is of order $1/g_\rho$, while that with the two NG bosons is of order~$g_\rho$. The diagram, and thus its contribution to the parameter
$\hat S$, is  naively of $O(g_\rho^0)$. There is however an enhanced contribution of $O(1/g_\rho^2)$
that comes from the kinematic region $s \sim M_\rho^2$  in the dispersive integral (\ref{eq:dispersionS}), where $M_\rho$ is the pole mass of the $\rho_L$.  
To see this, notice that the small $g_\rho$ limit coincides with a narrow-width expansion.
The Breit-Wigner function that follows from the square of the $\rho_L$ propagator can be thus expanded as
\begin{equation} \label{eq:BWexpansion}
\frac{\Gamma_\rho}{(s-M_\rho^2)^2 + M_\rho^2 \Gamma_\rho^2} = \frac{\pi}{M_\rho} \delta(s - M_\rho^2) + O(g_\rho^2) \, ,
\end{equation}
where $\Gamma_\rho$ is the decay width of the $\rho_L$. 
The left-hand side is of $O(g_\rho^2)$ for $s$ away from $M_\rho^2$, but the delta-function term in the right-hand side is of $O(g_\rho^0)$.
The contribution to the dispersive integral at the $\rho_L$ peak is thus enhanced compared to the naive counting. As a consequence, the leading
contribution to the $\hat S$ parameter from the $\pi\pi$ final state is of order $1/g_\rho^2$, and in fact  corresponds to the tree-level correction
of the diagrammatic calculation.

Loosely speaking, we can say that whenever the $\rho_L$ goes ``on shell'', the order in powers of $g_\rho$ is lowered by two units.
This has two consequences. The first is that the leading contribution from the $3\pi$ and $4\pi$  states can be captured by replacing them, respectively,
with  the states $\pi\rho_L$ and $\rho_L\rho_L$ obtained by treating the $\rho_L$  as an asymptotic state.
This approximation is sufficient to extract $\hat S$ at $O(g_\rho^0)$ and simplifies considerably the calculation.
The second consequence is that, in the calculation of the $\pi\pi$ contribution, 1-loop corrections to the vertices and to the $\rho_L$ propagator
should be included for $s \simeq M_\rho^2$, as they contribute at $O(g_\rho^0)$. In other words, 1-loop corrections to the spectral functions need to be 
retained (only) near the $\rho_L$ peak.

The Feynman diagrams relative to the calculation of the spectral functions $\tilde\rho_{LL}, \tilde\rho_{RR}$ and $\tilde\rho_{BB}$ are shown in Fig.~\ref{fig:diagrams} 
in terms of the relevant final states $\pi\pi$, $\rho_L\rho_L$ and $\pi \rho_L$.
We work in the unitary gauge for  $\rho_L$, choosing dimensional regularization and 
an on-shell minimal subtraction scheme~\cite{Contino:2015mha} to remove the divergences of the 1-loop contributions.
While the calculation of $\tilde\rho_{RR}$ and $\tilde\rho_{BB}$ is straightforward, it is worth discussing in some detail how the 1-loop corrections have been included 
in  $\tilde\rho_{LL}$.
As already stressed, we need to consider 1-loop effects only at the $\rho_L$ peak, for $s \sim M_\rho^2$. The first and third diagrams in the first row
of Fig.~\ref{fig:diagrams} can thus be evaluated at tree level. The second diagram gets 1-loop corrections in the vertex with the current (light blue blob with a cross),
the $\rho_L$ propagator (dark blue box) and the $\rho_L \pi\pi$ vertex (light blue blob).
By decomposing each of these three terms into a longitudinal and a transverse part, the contribution of the diagram to the matrix element of the current between 
the vacuum and  two NG bosons can be written as:
\begin{equation} \label{eq:matrixelement}
\begin{split}
\langle 0| \tilde J_\mu^{a_L} | \pi^k(p_1)\pi^l(p_2)\rangle \big|_\rho = 
& \, \delta^{a_L i} \left( \Pi_{J\rho}(q^2) P_{T\,\mu\alpha} + \bar{\Pi}_{J\rho}(q^2) P_{L\,\mu\alpha} \right) \\ 
& \times \delta^{ij} \left(  G(q^2)  P_T^{\alpha\beta} +  \bar G(q^2) P_L^{\alpha\beta}  \right) \\
& \times  \frac{1}{2} \epsilon^{jkl} \left[   (p_1 - p_2)_\beta V(q^2) +  q_\beta \bar{V}(q^2) \right]\, ,
\end{split}
\end{equation}
where $P^{\mu\nu}_T = (\eta^{\mu\nu} - q^\mu q^\nu/q^2)$, $P_L^{\mu\nu} = q^\mu q^\nu/q^2$ and $q = p_1 + p_2$.
The spectral function $\tilde\rho_{LL}$ is extracted by squaring this matrix element, integrating over the two-particle phase space and
finally projecting over the transverse part (see Eq.~(\ref{eq:defspectral})). The expression of the longitudinal terms in Eq.~(\ref{eq:matrixelement}) is thus not
relevant, as they do not enter the final result. For the transverse terms we use the following approximate expressions, 
\begin{align}
\label{eq:PiJrho}
\Pi_{J\rho}(q^2) & = \frac{m_\rho^2}{g_\rho} - 2 \alpha_2 g_\rho q^2 + g_\rho \Pi^{(1L)}_{J\rho}, \\[0.1cm]
G(q^2) &= \frac{\tilde{Z}_\rho}{q^2 - M_\rho^2 + i M_\rho \Gamma_\rho}, \\[0.1cm]
V(q^2) &=  \tilde{Z}_\rho^{-1/2} \left(96\pi \frac{\Gamma_\rho}{M_\rho} \right)^{1/2}\, , 
\end{align}
where the 1-loop parts have been evaluated at $q^2 = M_\rho^2$.
The quantity $\Pi^{(1L)}_{J\rho}$ encodes the pure 1-loop correction from NG bosons to the current-$\rho_L$ mixing. 
For the propagator $G(q^2)$ we make use of its resummed expression near the $\rho_L$ pole in terms of the pole mass $M_\rho$, total decay width $\Gamma_\rho$
and  pole residue $\tilde Z_\rho$.  Finally the vertex $V(q^2)$ is expressed in terms of the decay width $\Gamma_\rho$.
We report the analytic formulas for $\Pi^{(1L)}_{J\rho}$, $M_\rho^2$, $\tilde Z_\rho$ and $\Gamma_\rho$ in Appendix~\ref{app:spectralfunctions}.
Notice that a tree-level expression for $\Gamma_\rho$ is sufficient to reach the $O(g_\rho^0)$ precision we are aiming for in the spectral function.
Adding the contribution of the first diagram in the first row of Fig.~\ref{fig:diagrams} and inserting the total matrix element
in Eq.~(\ref{eq:defspectral}), one finds the following result for the spectral function
\begin{equation} \label{eq:anatomyrhoLL}
\tilde\rho_{LL}^{(\pi\pi)}(q^2) = \tilde\rho_{RR}(q^2) \times \big| 1 - a_\rho^2 +  \Pi_{J\rho}(q^2) G(q^2) V(q^2) \big|^2\, ,
\end{equation}
where $\tilde\rho_{RR}$ is given in Eq.~(\ref{eq:rhoRR}).
Away from the $\rho_L$ peak the 1-loop corrections can be neglected, and the second term in the absolute value in Eq.~(\ref{eq:anatomyrhoLL}) is of order $g_\rho^0$, 
like the first one. At the peak, on the other hand, this second term develops an $O(1/g_\rho^2)$ contribution. This can be identified by using 
Eq.~(\ref{eq:BWexpansion}) to expand $\tilde\rho_{LL}^{(\pi\pi)}(s)$  as a distribution. One has:
\begin{equation} \label{eq:expansionLL}
\tilde\rho_{LL}^{(\pi\pi)}(s) = Z_L M_\rho^2 \, \delta(s-M_\rho^2) + f_{LL}(s)\, .
\end{equation}
Here $Z_L$ is the pole residue of the two-point current correlator:
\begin{equation}
Z_{L}=  \left(\frac{1}{g_{\rho}}-2\alpha_{2}g_{\rho}\right)^{2} -\frac{2a_{\rho}^{4}-4a_{\rho}^{2}+85}{96\pi^{2}}\log\frac{\mu}{m_{\rho}} 
                      - \frac{10 a_\rho^4 - 32 a_\rho^2 + 1289 - 231 \pi\sqrt{3}}{576 \pi^2}\, .
\end{equation}
It is of order $1/g_\rho^2$ and, being an observable,  is RG invariant. The function $f_{LL}$ denotes instead the $O(g_\rho^0)$ continuum 
(which receives a contribution from both the NG bosons and the $\rho_L$).

\begin{figure}[t]
\begin{center}
\includegraphics[width=0.60\textwidth]{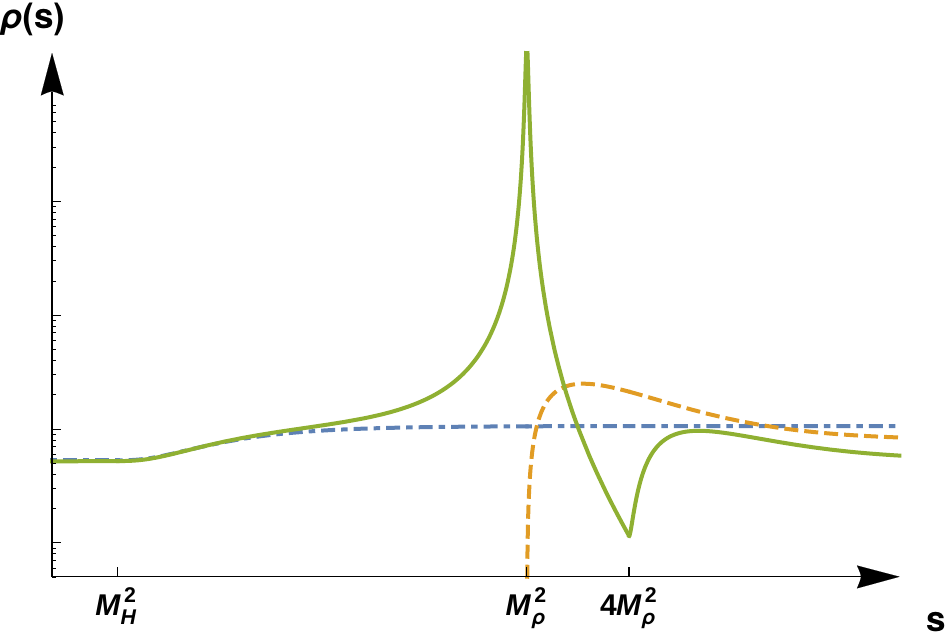}
\end{center}
\caption{\small Plot of the spectral functions $\tilde\rho_{LL}$ (continuous green curve), $\tilde\rho_{RR}$ (dot-dashed blue curve) and $\tilde\rho_{BB}$ (dashed orange curve), 
computed at $O(g_\rho^0)$ for the following  choice of parameters: $m_\rho(m_\rho) = 2\,$TeV, $g_\rho(m_\rho)=3$, $a_\rho = 1$ and $\alpha_2(m_\rho)=0$.
The kink of $\tilde\rho_{LL}$ at $s = 4 M_\rho^2$ is due to the onset of the contribution of the $\rho_L\rho_L$ intermediate state.
The scale is logarithmic on both axes.
}
\label{fig:spectralfunctions}
\end{figure}

The analytic expressions of the spectral functions are reported in Appendix~\ref{app:spectralfunctions}. Their plot (in $D=4$ dimensions) is shown in 
Fig.~\ref{fig:spectralfunctions} 
for the following benchmark choice of parameters: $m_\rho(m_\rho) = 2\,$TeV, $g_\rho(m_\rho)=3$, $a_\rho = 1$ and $\alpha_2(m_\rho)=0$ (here $m_\rho(\mu)$, $g_\rho(\mu)$
and $\alpha_2(\mu)$ are the running parameters, see Ref.~\cite{Contino:2015mha}).~\footnote{We have checked that setting $\alpha_2$ to a value of order $1/16\pi^2$ 
at the scale $m_\rho$, as obtained if $\alpha_2 =0$ at the cutoff scale, does not  change qualitatively the plot. Notice  that the running of $a_\rho$ arises at the
two-loop level~\cite{Contino:2015mha} and can be thus neglected. }
%
%
One can notice the following. The functions $\tilde\rho_{LL}(s)$ and $\tilde\rho_{RR}(s)$ 
become constant  and equal for $s\to 0$ (in $D=4$).
This constant tail corresponds to the NG boson contribution to the spectral functions; it  gives rise to the IR logarithmic singularity in the $\hat S$ parameter
that is eventually canceled by the subtraction in Eq.~(\ref{eq:dispersionS}).
Having set $\alpha_2 =0$, the spectral functions tend to a constant also for $s\to \infty$. This gives rise to a UV logarithmic divergence
in the spectral integral for $\hat S$ which can be regulated by extending the theory to $D$ dimensions (Notice that one should consistently
extend both the spectral functions and also the subtraction term in Eq.~(\ref{eq:dispersionS})).
The divergence is canceled by the local counterterm generated by the operator $O_3^+ = \Tr[(E_{\mu\nu}^L)^2 + (E_{\mu\nu}^R)^2]$.
The correlator~$\Pi_1$ thus obeys a dispersion relation of the form~(\ref{eq:dispersionMinfinite}),
\begin{equation}
\label{eq:dispersionforPi1tilde}
\Pi_{1}(q^2) = \tilde\Pi_{1}(0) + q^2 \int_0^{\infty} \!\! ds\; \frac{\tilde\rho_{LL}(s) + \tilde\rho_{RR}(s) -2 \tilde\rho_{BB}(s)}{s-q^2} + q^2 (C_{1} - 8 c_3^+) + \dots  \, ,
\end{equation}
where  $C_1 \equiv \underset{q^2\to \infty}{\lim}  [\tilde\Pi_1(q^2)/q^2]$, $c_3^+$ is the coefficient of $O_3^+$,
and the dots indicate local terms with higher powers of $q^2$.
For $\alpha_2 =0$ the contribution from the integral on the circle vanishes, $C_1 =0$,  when extending the theory to $D$ dimensions.
For non-vanishing $\alpha_2$, on the other hand, $\tilde\Pi_1(q^2)$ grows like $q^2$ in any dimension (as a consequence of its tree-level behavior) 
and one finds $C_1 =  -4\alpha_2^2 g_\rho^2$.

Using the expressions of the spectral functions 
we can derive our final expression for~$\hat S$. 
We find:
\begin{equation}
\label{eq:finalS}
\begin{split}
\hat{S} =& \, \frac{g^2}{4g_\rho^2} \sin^2\!\theta \left(1 - 2\alpha_2 g_\rho^2\right)^2 + \frac{g^2}{96\pi^2} \sin^2\!\theta 
                  \left( \log\frac{\mu}{m_h} + \frac{5}{12} \right) \\
              &- \frac{g^2}{96\pi^2}\sin^2\!\theta \left[ \frac{3}{4}\left(a_\rho^2 + 28\right) \log\frac{\mu}{m_\rho} + 1 + \frac{41}{16}a_\rho^2 \right]
                + g^2\sin^2\!\theta\left(-2 c_3^+(\mu) + \frac{C_1}{4} \right)\, .
\end{split}
\end{equation}
Notice that the term proportional to $C_1$ cancels the $\alpha_2^2$ part in the first term.

The parameters $W$ and $Y$ obey the same dispersion relations of the full theory, Eqs.~(\ref{eq:dispersionW}) and~(\ref{eq:dispersionY}), 
with $\rho_{ij}$ replaced by the  spectral functions of the effective theory $\tilde\rho_{ij}$. All contributions from the integrals on the circle,  in this case,
can be made to vanish through dimensional continuation.
The contact terms to be added in the effective theory are generated by the  operators $O_{2W} = (\nabla^\mu E^L_{\mu\nu})^2$ and $O_{2B} = (\nabla^\mu E^R_{\mu\nu})^2$.
Their contribution is naively of  $O[(m_W^2/m_\rho^2)(g^2/16\pi^2)]$, i.e. of higher order in our approximation, and will be thus neglected.
Furthermore, since we are interested  in the leading
correction of $O[(m_W^2/m_\rho^2)(g^2/g_\rho^2)]$ from the $\rho_L$, the integral in Eq.~(\ref{eq:dispersionW})
can be computed by retaining only the delta function in the expansion of $\tilde\rho_{LL}$ in Eq.~(\ref{eq:expansionLL}) (while that in Eq.~(\ref{eq:dispersionY}) is
negligible). 
We thus find:~\footnote{The $O(m_W^4/m_\rho^4)$ terms of footnote~\ref{fot:neglectedterms} give the additional corrections
\begin{equation}
\delta W =  \frac{m_W^2}{m_\rho^2} \, \frac{g^2}{g_\rho^2} \left( 1 - 2 \alpha_2 g_\rho^2 \right)^2   \left(  \cos^4\!\frac{\theta}{2} -1 \right) \, , \qquad\quad
\delta Y  =  \frac{m_W^2}{m_\rho^2} \, \frac{g'^2}{g_\rho^2} \left( 1 - 2 \alpha_2 g_\rho^2 \right)^2  \sin^4\!\frac{\theta}{2}\, ,
\end{equation}
which also come from  the delta function in the expansion of $\tilde\rho_{LL}$.
}
\begin{align}
\label{eq:finalW}
\begin{split}
W  = &\, \frac{g^2}{96\pi^2} c_W^2 \sin^2\!\theta \left[  \frac{9 x_h^2 +12 x_h}{2 (x_h -1)^5} \log x_h  - \frac{x_h^3 + x_h^2 + 73 x_h +9}{8(x_h -1)^4} \right] \\
  & \, + \frac{m_W^2}{m_\rho^2} \, \frac{g^2}{g_\rho^2} \left( 1 - 2 \alpha_2 g_\rho^2 \right)^2  \, , 
\end{split}
\\[0.5cm]
\label{eq:finalY}
Y  = & \, \frac{g'^2}{96\pi^2} c_W^2 \sin^2\!\theta \left[  \frac{9 x_h^2 +12 x_h}{2 (x_h -1)^5} \log x_h  - \frac{x_h^3 + x_h^2 + 73 x_h +9}{8(x_h -1)^4}\right]\, .
\end{align}

Using Eqs.~(\ref{eq:finalS}),~(\ref{eq:finalW}) and~(\ref{eq:finalY}), together with Eq.~(\ref{eq:longdistance}), we obtain our final formula for $\Delta\eps_3$:
\begin{equation} 
\label{eq:finaldeps3}
\begin{split}
\Delta\epsilon_3 = &\, \frac{g^2}{4g_\rho^2} \sin^2\!\theta \left(1 - 4\alpha_2g_\rho^2\right) 
                                     + \frac{m_W^2}{m_\rho^2} \, \frac{g^2}{g_\rho^2}  \left(1 - 2\alpha_2g_\rho^2\right)^2 \\
                              &   -2 g^2\sin^2\!\theta \, c_3^+(\mu)  + \frac{g^2}{96\pi^2} \sin^2\!\theta \left( \log\frac{\mu}{m_Z} + f_3(h) \right) \\
                              &- \frac{g^2}{96\pi^2}\sin^2\!\theta \left[ \frac{3}{4}\left(a_\rho^2 + 28\right)\log\frac{\mu}{m_\rho} + 1 + \frac{41}{16}a_\rho^2 \right]\, .
\end{split}
\end{equation}
%

\section{Discussion and conclusions}
\label{sec:discussion} 

Equation~(\ref{eq:finaldeps3}) coincides with the result that we obtained in Ref.~\cite{Contino:2015mha} through a 1-loop diagrammatic calculation of 
$\Delta\eps_3$.~\footnote{The $O[(m_W^2/m_\rho^2)(g^2/g_\rho^2)]$ contribution from $W$ and $Y$ was neglected in Ref.~\cite{Contino:2015mha}, 
see Eq.~(4.47) therein.} It shows that at tree level (i.e. at $O(1/g_\rho^2)$)
the sign of $\Delta\eps_3$, as well as that of $\hat S$ in Eq.~(\ref{eq:finalS}), is controlled by $\alpha_2$ and 
is not necessarily positive. This was considered problematic by Rychkov and Orgogozo  in their analysis of Ref.~\cite{Orgogozo:2012ct}, based on the expectation 
that $\hat S$ should be positive if obtained through a dispersion  relation where the leading contribution arises from the (positive definite) spectral function $\rho_{LL}$.
They suggested that the positivity of $\hat S$ is in fact restored once the correct asymptotic behavior in the deep Euclidean ($q^2 \to - \infty$) implied by the OPE 
is enforced on the expressions of the  two-point current correlators computed in the effective theory. In particular,  one 
expects that $\Pi_1(q^2) \sim (-q^2)^{1-\Delta_1/2}$ for $q^2\to -\infty$, where $\Delta_1 \geq 1$ is the scaling dimension of the first scalar operator contributing to its OPE.
If this condition is enforced on Eq.~(\ref{eq:dispersionforPi1tilde}) by neglecting the higher-derivative terms denoted by the dots, one obtains 
$c_3^+ = C_1/8 = - \alpha_2^2 g_\rho^2/2$, where from now on we focus on the tree-level contribution neglecting the $O(1/16\pi^2)$ radiative corrections.
This relation implies that the last term of Eq.~(\ref{eq:finalS}) identically vanishes,
giving the positive definite expression derived in 
Ref.~\cite{Orgogozo:2012ct}:  $\hat S = (g^2 \sin^2\!\theta/4 g_\rho^2)(1-2\alpha_2 g_\rho^2)^2$.  Now, the higher-derivative terms in Eq.~(\ref{eq:dispersionforPi1tilde}) 
are suppressed by corresponding powers of the cutoff scale $\Lambda$.  As such they become important at 
energies $E \sim \Lambda$. Neglecting them when enforcing the asymptotic behavior is in fact equivalent to requiring that this latter is attained at energies
$E \sim M_\rho$ through the exchange of the $\rho_L$, while the cutoff states have no effect.
In this sense, the correction coming from $c_3^+$ should be regarded as characterizing part of the $\rho_L$ contribution rather than encoding
the effect of the cutoff states. Requiring that the asymptotic behavior  be obtained at the scale $M_\rho$, as effectively done in Ref.~\cite{Orgogozo:2012ct},
thus leads to a positive $\hat S$.

There is, on the other hand, the possibility that the correct asymptotic behavior is recovered only at energies $E \sim \Lambda$ as the effect of the 
higher-derivative terms.
That is to say, it can be enforced by the exchange of the  cutoff states rather than by the lighter resonance~$\rho_L$. 
In this case it is reasonable to assume $c_3^+ <  1/g_\rho^2$, as suggested by its naive estimate, so that
$\hat S = (g^2 \sin^2\!\theta/4 g_\rho^2)(1-4\alpha_2 g_\rho^2)$ up to smaller corrections. This expression is not definite positive, as previously noticed. 
It is a result consistent with the properties of the underlying strong dynamics and in fact  plausible to some degree.
Indeed,  the  behavior of the correlators in the deep Euclidean could be determined by the dynamics at or beyond the cutoff scale, 
while the $\hat S$  parameter  is saturated in the infrared and as such gets its leading contribution from the lightest modes.
A simple model with three spin-1 resonances is discussed in Appendix~\ref{app:example} which illustrates this possibility with an explicit example.

The tree-level value of the $\hat S$ parameter can then be tuned to be small or may even become negative for  $\alpha_2$ of order
$1/g_\rho^2$. While such large values are not expected  from a naive estimate if $\alpha_2$ is generated by the physics at the cutoff scale
(in this case one would expect $\alpha_2 \sim f^2/\Lambda^2$ or smaller), they are consistent with the request of the absence of a ghost in the low-energy
theory~\cite{Contino:2011np}. Having $\alpha_2 \sim 1/g_\rho^2$, on the other hand, affects the naive estimate of~$c_3^+$. For non-vanishing $\alpha_2$,
the 1-loop correction to $\tilde\Pi_1'(0)$ is quadratically divergent, which implies $c_3^+(\Lambda) \sim  (\Lambda^2/m_\rho^2) (\alpha_2^2 g_\rho^4)/16\pi^2$.
For $\alpha_2 \sim 1/g_\rho^2$ and setting $\Lambda = g_* f$ one has $c_3^+(\Lambda) \sim g_*^2/(16\pi^2 g_\rho^2)$.
This can be as large as the tree-level contribution from the $\rho_L$ exchange if $g_* \sim 4\pi$.  Such enhancement of the 1-loop
contribution from the cutoff dynamics originates from the increased coupling strength through which the transverse gauge fields
interact with the composite states.
In particular, the $\pi\pi W\rho_L$ vertex gets an energy-growing contribution of order $g g_\rho (\alpha_2 g_\rho^2) E^2/m_\rho^2$.
For $\alpha_2 \sim 1/g_\rho^2$, this translates into a coupling strength squared of order $g g_* (g_*/g_\rho)$ at the cutoff scale, which is a factor 
$(g_*/g_\rho)$ stronger than the naive estimate based on the Partial UV Completion (PUVC) criterion~\cite{Contino:2011np}. 
This is precisely the enhancement factor appearing in the estimate of $c_3^+$.
We thus conclude that while  for $\alpha_2 \sim1/g_\rho^2$ it is possible to make  the tree-level value of $\hat S$ small or even negative, this is at the price
of increasing the naive size of the unknown contribution from the cutoff states. 
Such a contribution becomes of order $1/g_\rho^2$ if $g_* \sim 4\pi$, making the $\hat S$ parameter in practice incalculable in the effective theory.

As a final remark we notice that when including the 1-loop corrections, the asymptotic behavior of the full theory is not attained at $M_\rho$ even for $\alpha_2 =0$.
In fact, one has $\tilde\Pi_1(q^2) \sim  q^2 \log (-q^2) (1-a_\rho^2)(5/2 - a_\rho^2)$ for $q^2 \to -\infty$ (in $D=4)$. 
Setting $a_\rho^2$ equal to 1 or $5/2$ (and $\alpha_2=0$) thus gives a model of the strong dynamics where the
asymptotic behavior of $\Pi_1$ is  enforced by the exchange of the $\rho_L$,
and the dispersive integral of the $\hat S$ parameter in the effective theory is convergent in $D=4$. 
In a low-energy theory with both $\rho_L$ and $\rho_R$, one has that $\tilde \Pi_1(q^2)/q^2$ vanishes at infinity 
for $a^2_{\rho_L} = a^2_{\rho_R} = 1/2$ or 3 (and $\alpha_{2L} = \alpha_{2R} =0$). 
The choice $a^2_{\rho_L} = a^2_{\rho_L} = 1/2$, in particular,  corresponds 
to a two-site model limit in which the global symmetry is enhanced to $SO(5)\times SO(5)\to SO(5)$~\cite{Contino:2015mha}. The finiteness of the $\hat S$ parameter
in this case follows as a consequence of the larger symmetry.~\cite{Panico:2011pw,Contino:2015mha}

\vspace{0.4cm}
In this paper we have derived dispersion relations for the electroweak oblique parameters in the context of $SO(5)/SO(4)$ composite Higgs theories.
We have distinguished between long- and short-distance contributions to $\eps_3$, 
and obtained a dispersion relation for each of the parameters $\hat S$, $W$ and $Y$ characterizing the short-distance part
(Eqs.(\ref{eq:dispersionS}),~(\ref{eq:dispersionW}) and~(\ref{eq:dispersionY})).
Our analysis generalizes the dispersion relation  written by Peskin and Takeuchi for the $S$ parameter in the case of Technicolor~\cite{Peskin:1991sw}.
We thus derived a dispersion relation for~$\eps_3$ (Eq.~(\ref{eq:dispersioneps3})), extending  the work of Rychkov and Orgogozo~\cite{Orgogozo:2012ct}.
Our formula (\ref{eq:dispersioneps3}) agrees with their result and further reduces the relative theoretical  uncertainty to order $m_h^2/m_*^2$, 
where $m_*$ is the mass scale of the resonances of the strong sector. This is to be compared with the $O(m_h/m_*)$ relative uncertainty of Ref.~\cite{Orgogozo:2012ct}.
We also discussed how the dispersion relations can be used and get modified in the  context of a low-energy effective description of the strong dynamics.
Making use of dimensional regularization we provided a definition of the otherwise divergent spectral integrals, pointing out the importance of the contribution
from the integral on the circle in the case in which the two-point correlators of the effective theory do not die off fast enough at infinity.
We utilized our formula to perform the dispersive calculation of $\eps_3$ at the 1-loop level in a theory with a \mbox{spin-1} resonance $\rho_L$. 
We pointed out that 1-loop corrections need to be retained only at the $\rho_L$ peak to obtain $\eps_3$ at the $O(g_\rho^0)$ level. This considerably simplified
our calculation and conveniently reproduced the result of the diagrammatic computation that we performed in Ref.~\cite{Contino:2015mha}.
The dispersive approach is particularly suitable to clarify the connection between the positivity of the $\hat S$ parameter and the 
UV behavior of two-point current correlators, as first suggested by Ref.~\cite{Orgogozo:2012ct}. We argued that if the behavior dictated by the OPE in the deep Euclidean 
is enforced at the scale $M_\rho$ through the exchange of the light resonances, then the $\hat S$ parameter is positive definite in agreement with the expectation 
of Ref.~\cite{Orgogozo:2012ct}. It is possible, on the other hand, that the UV behavior is recovered only at the cutoff scale as an effect of the heavier resonances, 
while the leading contribution to the $\hat S$ parameter is still saturated by the lowest lying modes. In this case $\hat S$ can be negative if the $\rho_L$ dynamics 
is characterized by a large kinetic mixing with the gauge fields of order $\alpha_2 \sim 1/g_\rho^2$.

\section*{Acknowledgments}

We would like to thank 
Marco Bochicchio,
Gino Isidori,
Agostino Patella,
Riccardo Rattazzi,
Massimo Testa and
Enrico Trincherini
for discussions, and  especially Slava Rychkov for important discussions and suggestions.
The work of R.C. was partly supported by the ERC Advanced Grant No.~267985 
\textit{Electroweak Symmetry Breaking, Flavour and Dark Matter: One Solution for Three Mysteries (DaMeSyFla)}.

\appendix

\section{Generalization to the case of strong dynamics with small $SO(5)$ breaking}
\label{app:SO5approximatecase} 

In deriving our dispersion relations we have assumed that the strong dynamics in isolation is $SO(5)$ symmetric.
It is conceivable, on the other hand, that the global symmetry is only approximate and that a small explicit breaking
arises internal to the strong dynamics. This is for example what happens in the Minimal Conformal Technicolor model of Ref.~\cite{Luty:2004ye},
where the small breaking arises from the techniquark mass terms.
Generalizing our procedure to such a scenario is straightforward. We will assume that an $SO(3)\times P_R$ subgroup of the strong
dynamics is unbroken, where $SO(3)$ is the custodial isospin and $P_R$ is the grading of the $SO(5)$ algebra under which the $SO(5)/SO(4)$ 
generators are odd. This allows for a Higgs boson potential, hence a Higgs mass, ensuring a correct phenomenology.
The definitions of the two-point correlators generalizing Eq.~(\ref{eq:2pointGF}) thus read:
\begin{equation} \label{eq:2pointGFmH}
\begin{split}
\langle J^{a_L}_\mu(q)  J_\nu^{b_L}(-q) \rangle  = &\, -i \delta^{a_L b_L} \left( \eta_{\mu\nu} \Pi_{LL}(q^2) - q_\mu q_\nu   \bar \Pi_{LL}(q^2) \right) \\[0.15cm]
\langle J^{a_R}_\mu(q)  J_\nu^{b_R}(-q) \rangle  = & \, -i \delta^{a_R b_R} \left( \eta_{\mu\nu} \Pi_{RR}(q^2) - q_\mu q_\nu   \bar \Pi_{RR}(q^2) \right) \\[0.15cm]
\langle J^{a_L}_\mu(q)  J_\nu^{b_R}(-q) \rangle  = & \, -i \delta^{a_L b_R} \left( \eta_{\mu\nu} \Pi_{LR}(q^2) - q_\mu q_\nu   \bar \Pi_{LR}(q^2) \right) \\[0.15cm]
\langle J^{\hat a}_\mu(q)  J_\nu^{\hat b}(-q) \rangle  = & \,  -i \delta^{\hat a \hat b}  \left( \eta_{\mu\nu} \Pi_{BB}(q^2) - q_\mu q_\nu  \bar \Pi_{BB}(q^2) \right) \\[0.05cm]
                                                                               & \, -i \delta^{\hat a 4} \delta^{\hat b 4}  \left( \eta_{\mu\nu} \Pi^{(4)}_{BB}(q^2) - q_\mu q_\nu  \bar \Pi^{(4)}_{BB}(q^2) \right) \, .
\end{split}
\end{equation}
Any two-point function with one $SO(5)/SO(4)$ and one $SO(4)$ current vanishes due to $P_R$ invariance.
As a consequence of the $SO(5)$ breaking, in particular, $\Pi_{LR}$ does not vanish and must be included in the definition of $\Pi_{3B}$ when deriving Eq.~(\ref{eq:Spi3B}):
\begin{equation}
\label{eq:rel3BmH}
\Pi_{3B}(q^2) \equiv \frac{1}{4} \sin^2\!\theta \left( \Pi_{LL}(q^2) + \Pi_{RR}(q^2) - 2 \Pi_{BB}(q^2) \right) + \frac{1}{2} \left( 1 + \cos^2\!\theta\right) \Pi_{LR}(q^2) \, .
\end{equation}
Since now the Higgs boson mass is non-vanishing, Eq.~(\ref{eq:Spi3B}) is free from IR singularities, which cancel when taking the difference with the SM.
It is still convenient, however, to add and subtract the contribution from the $SO(5)/SO(4)$ linear model, as was done in the text.
A first motivation to do so is that the $SO(5)$ breaking internal to the strong dynamics only partly accounts for the Higgs mass; an important (if not dominant)
contribution comes from the coupling to the elementary top quark, which is not included.
The second motivation is that subtracting the $SO(5)/SO(4)$ linear model allows one to isolate the Higgs chiral logarithm, so that the final dispersive integral
encodes the contribution from the heavy resonances only.
By performing the subtraction as explained in the text, the result that follows coincides with the massless case. That is, Eq.~(\ref{eq:Sfinal}) is valid also in the massive case, 
with $\Pi_{3B}$ defined as in Eq.~(\ref{eq:rel3B}). 
This is because the only unsuppressed contribution to $\Pi_{LR}$ comes from the NG bosons and cancels out when subtracting the $SO(5)/SO(4)$ linear model.
Although Eq.~(\ref{eq:Sfinal}) is formally unchanged, $\Pi_{3B}^{LSO5\, \prime}(0)$ in parenthesis must be evaluated by setting the Higgs mass to the same value 
$m_{0h}$ generated by the strong dynamics. The dispersion relation generalizing Eq.~(\ref{eq:dispersionS}) reads
\begin{equation} \label{eq:dispersionSmH}
\begin{split}
\hat S = \frac{g^2}{4}  \sin^2\!\theta \int_0^\infty \! \frac{ds}{s} \, \Bigg\{ 
 & \left(  \rho_{LL}(s) +  \rho_{RR}(s) - 2 \rho_{BB}(s) \right)   \\
 & - \frac{1}{48\pi^2} \Bigg[  \, \frac{1}{2} + \frac{1}{2} \left(  1 - \frac{m_{0h}^2}{s} \right)^3 \theta(s - m_{0h}^2)  \\
 &  \phantom{- \frac{1}{48\pi^2} \Bigg[  \,} \, - \left(  1 - \frac{m_\eta^2}{s} \right)^3 \theta(s - m_\eta^2) \Bigg]
\Bigg\} + \delta \hat S_{LSO5} + \delta\hat S_{Zh} \, ,
\end{split}
\end{equation}
where $\delta \hat S_{LSO5}$ is still defined by Eq.~(\ref{eq:deltaLSO5}) and computed at the physical Higgs mass.
Similarly, the dispersion relations for $W$ and $Y$ are:
\begin{align}
\label{eq:dispersionWmH}
\begin{split}
W = & - m_W^2 g^2  \int^\infty_0 \! \frac{ds}{s^2}\, \left\{ \rho_{LL}(s) - \frac{1}{192\pi^2} \left[  1 + \left(  1 - \frac{m_{0h}^2}{s} \right)^3 \theta(s - m_{0h}^2)  \right]\right\}\\
       & + \frac{g^2}{96\pi^2} \frac{c_W^2}{8 x_h} \sin^2\!\theta  + \delta W_{Zh} 
\end{split} \\[0.3cm]
\label{eq:dispersionYmH}
\begin{split}
Y = & - m_W^2 g'^2  \int^\infty_0 \! \frac{ds}{s^2}\, \left\{ \rho_{RR}(s) - \frac{1}{192\pi^2} \left[  1 + \left(  1 - \frac{m_{0h}^2}{s} \right)^3 \theta(s - m_{0h}^2)  \right]\right\}\\
       & + \frac{g'^2}{96\pi^2} \frac{c_W^2}{8 x_h} \sin^2\!\theta  + \delta Y_{Zh} \, .
\end{split} 
\end{align}
The  formula for $\Delta\eps_3$ finally reads:
\begin{equation}
\label{eq:dispersioneps3mH}
\begin{split}
\Delta\eps_3 = & \, \frac{g^2}{96\pi^2} \sin^2\!\theta \left(f_3(x_h) - \frac{1}{8x_h} + \frac{\log x_h}{2}  - \frac{5}{12}  + \log\frac{m_\eta}{m_h}  \right)  \\
 & + \frac{g^2}{4}  \sin^2\!\theta \int_0^\infty \! \frac{ds}{s} \, \Bigg\{ 
     \rho_{LL}(s) +  \rho_{RR}(s) - 2 \rho_{BB}(s)   \\
 & \phantom{+ \frac{g^2}{4}  \sin^2\!\theta \int_0^\infty \! \frac{ds}{s} \, \Bigg\{ \, } 
    - \frac{1}{96\pi^2} \Bigg[  \, \frac{1}{2} + \frac{1}{2} \left(  1 - \frac{m_{0h}^2}{s} \right)^3 \theta(s - m_{0h}^2) \\
  & \hspace{5.3cm} - \left(  1 - \frac{m_\eta^2}{s} \right)^3 \theta(s - m_\eta^2) \Bigg] \Bigg\}  \\
 & + m_W^2  \int^\infty_0 \! \frac{ds}{s^2}\, \Bigg\{  g^2  \rho_{LL}(s) + g'^2  \rho_{RR}(s) \\
  &  \hspace{2.9cm}   -  \frac{g^2+g'^2}{192\pi^2}  \Bigg[  1+ \left(  1 - \frac{m_{0h}^2}{s} \right)^3 \theta(s - m_{0h}^2) \Bigg] \Bigg\} \, .
\end{split}
\end{equation}
Notice that the dependence on $m_{0h}$ in Eqs.~(\ref{eq:dispersionSmH})-(\ref{eq:dispersioneps3mH}) cancels out up to negligible terms with relative suppression 
of order $m_{0h}^2/m_*^2$.

\section{Spectral functions and useful formulas}
\label{app:spectralfunctions} 

We report here the expressions of the spectral functions computed in the low-energy effective theory in $D$ dimensions, 
which can be used to perform the dispersive integrals using dimensional regularization. For convenience they are given for a finite Higgs mass $m_h$, 
so that one should set $m_h =0$ in evaluating the integrals of Eqs.~(\ref{eq:dispersionS}),~(\ref{eq:dispersionW}),~(\ref{eq:dispersionY}) and (\ref{eq:dispersioneps3}).
The $LL$ and $RR$ spectral functions are computed by introducing a small mass $\lambda$ for the three $SO(4)/SO(3)$ NG bosons which acts as an IR regulator
when considering their individual contribution to the dispersive integrals. Notice, on the other hand, that the linear combination of spectral functions appearing
in Eqs.~(\ref{eq:dispersionS}),~(\ref{eq:dispersionW}),~(\ref{eq:dispersionY}) and (\ref{eq:dispersioneps3}) is free from IR divergences, and that one should
set $\lambda =0$ when evaluating them.

The function $\tilde\rho_{RR}$ receives a contribution from the intermediate states $\chi\chi$ and $\chi h$, where $\chi^{1,2,3} \equiv \pi^{1,2,3}$ and $h = \pi^4$.
We find:
\begin{align}
\label{eq:rhoRR}
\tilde\rho_{RR}(q^2) & = \tilde\rho^{(\chi\chi)}_{RR}(q^2)  + \tilde\rho^{(\chi h)}_{RR}(q^2)\, , \\[0.7cm]
\tilde\rho^{(\chi\chi)}_{RR}(q^2) & = \frac{\mu^{4-D}}{\pi^{(D-1)/2}\, 4^D \,\Gamma\!\left(\frac{D+1}{2}\right)} \left(1 - 4\frac{\lambda^2}{q^2}\right)^{(D-1)/2} (q^2)^{(D-4)/2}
                                     \,\theta\!\left(q^2-4\lambda^2\right)\, ,
\\[0.3cm]
\tilde\rho^{(\chi h)}_{RR}(q^2) & = \frac{\mu^{4-D}}{\pi^{(D-1)/2}\, 4^D \,\Gamma\!\left(\frac{D+1}{2}\right)} \left(1 - \frac{m_h^2}{q^2}\right)^{D-1} (q^2)^{(D-4)/2}
                                      \,\theta\!\left(q^2-m_h^2\right)\, .
\end{align}
The intermediate states contributing to $\tilde\rho_{LL}$ are $\pi\pi$ and $\rho\rho$. We have:
\begin{align}
\tilde\rho_{LL}(q^2) = & \, \tilde\rho^{(\pi\pi)}_{LL}(q^2)  + \tilde\rho^{(\rho\rho)}_{LL}(q^2)\, , \\[0.7cm]
\begin{split}
\tilde\rho^{(\rho\rho)}_{LL}(q^2) = & \, \frac{\mu^{4-D}}{\pi^{(D-1)/2}\, 4^D\, \Gamma\!\left(\frac{D+1}{2}\right)} \, 
                                                         \frac{q^4+20q^2m_\rho^2 +12m_\rho^4}{\left(q^2-m_\rho^2\right)^2}  \\[0.3cm]
                                                 &\times \left(1-4\frac{m_\rho^2}{q^2}\right)^{3/2} \left(q^2-4m_\rho^2\right)^{(D-4)/2} \,\theta\!\left(q^2 - 4M_\rho^2\right)\, .
\end{split}
\end{align}
where $\tilde\rho^{(\pi\pi)}_{LL}(q^2)$ is given by Eq.~(\ref{eq:anatomyrhoLL}).
Finally, the only contribution to $\tilde\rho_{BB}$ is from the intermediate state $\rho\pi$:
\begin{equation}
\begin{split}
\tilde\rho_{BB}(q^2) = &\, \frac{3\mu^{4-D}}{2\pi^{(D-1)/2}\, 4^D\, \Gamma\!\left(\frac{D+1}{2}\right)} \, a_\rho^2 
                                            \left(1 + 10 \frac{m_\rho^2}{q^2} + \frac{m_\rho^4}{q^4}\right)  \\[0.2cm]
                                        & \times \left(1 - \frac{m_\rho^2}{q^2}\right)^{D-3}  (q^2)^{(D-4)/2} \,\theta\!\left(q^2 - M_\rho^2\right)\, .
\end{split}
\end{equation}
Notice that for simplicity  the contribution of $\alpha_2$ has been included only in $\tilde\rho_{LL}^{(\pi\pi)}$, see Eq.~(\ref{eq:PiJrho}), and omitted in $\tilde\rho_{LL}^{(\rho\rho)}$
and $\tilde\rho_{BB}$. This corresponds to including $\alpha_2$ only at the tree level in a diagrammatic calculation, see Ref.~\cite{Contino:2015mha}.


For completeness, we also report the expression for the $\rho_L$ pole mass squared $M^2_\rho$, the pole residue $\tilde Z_\rho$, the decay width $\Gamma_\rho$ 
(tree-level expression), and the 1-loop vertex correction $\Pi^{(1L)}_{J\rho}$ used in Section~\ref{sec:calculation}:
\begin{align}
M_\rho^2 & =  m_\rho^2 - m_\rho^2 \, \frac{g_\rho^2}{96\pi^2}  
                  \left[  \left(2a_\rho^4 - 69\right) \log\frac{\mu}{m_\rho} + \frac{8}{3}a_\rho^4 - 103 + \frac{33\sqrt{3}}{2}\pi \right]\, , \\
\tilde Z_\rho & = 1 - \frac{g_\rho^2}{96\pi^2}  \left[  \left(2a_\rho^4 - 53\right) \log\frac{\mu}{m_\rho} + \frac{5}{3}a_\rho^4 - \frac{53}{6} - \frac{11\sqrt{3}}{2}\pi \right]\, , \\
\Gamma_\rho & = \frac{g_\rho^2a_\rho^4 }{96\pi}\,  m_\rho  \, , \\
\Pi^{(1L)}_{J\rho} & = -\frac{1}{48\pi^2} m_\rho^2 a_\rho^2 \left(a_\rho^2-1\right) \left(\log\frac{\mu}{m_\rho}+\frac{4}{3}+\frac{i}{2}\pi\right) \, .
\end{align}

\section{Model with asymptotic behavior recovered at the cutoff scale}
\label{app:example} 

A simple model can be constructed which illustrates the possibility that the asymptotic behavior of the correlator $\Pi_1(q^2)$ is enforced by the
exchange of the states at the cutoff scale, while the leading contribution to the $\hat S$ parameter is dominated by the lighter resonances.

Consider a low-energy theory with three spin-1 resonances transforming,
respectively, as a $(3,1)$ (the $\rho_L$), a $(1,3)$ ($\rho_R$) and a $(2,2)$ ($\rho_B$) of $SU(2)_L \times SU(2)_R$.
We will assume for the moment that  their  masses are all of the same order and accidentally (much) lighter than the cutoff scale.
The Lagrangian characterizing the $\rho_R$ is defined in Ref.~\cite{Contino:2015mha} and can be obtained from that of the $\rho_L$ through an obvious 
$L \leftrightarrow R$ exchange. The $\rho_B$ is instead described by
\begin{equation}
\label{eq:LrhoB}
{\cal L}^{(\rho_B)} = - \frac{1}{4g_{\rho_B}^2} \, \Tr\!\left[ \rho_{\mu\nu}^B \rho^{B\, \mu\nu} \right] - \frac{m_{\rho_B}^2}{2 g_{\rho_B}^2} \, \Tr\!\left[ \rho_\mu^B \rho^{B\, \mu}\right] +
                            \alpha_{2B} \, \Tr\!\left[ \rho_{\mu\nu}^B f^{-\, \mu\nu} \right]\, ,
\end{equation}
where $\rho_{\mu\nu}^B \equiv \nabla_\mu \rho_\nu^B - \nabla_\nu \rho_\mu^B$ and $f^-_{\mu\nu}$ is the component of the dressed field strength along the broken 
$SO(5)/SO(4)$ generators~\cite{Contino:2011np}.
A simple calculation shows that in the deep Euclidean  $\tilde\Pi_{LL}(q^2)/q^2 \simeq 4 \alpha_{2L}^2 g_{\rho_L}^2$, $\tilde\Pi_{RR}(q^2)/q^2 \simeq 4 \alpha_{2R}^2 g_{\rho_R}^2$ 
and $\tilde\Pi_{BB}(q^2)/q^2 \simeq 4 \alpha_{2B}^2 g_{\rho_B}^2$, where the $L,R,B$ subindices are used to denote the parameters of the corresponding resonances. 
The  asymptotic behavior $\Pi_{LL}(q^2) \sim \Pi_{RR}(q^2) \sim \Pi_{BB}(q^2) \sim \gamma \, q^2$, where $\gamma$
is a constant proportional to the central charge of the OPE, is thus reproduced by the correlators in the effective theory if
\begin{equation}
\label{eq:condition}
\alpha_{2L}^2 g_{\rho_L}^2 = \alpha_{2R}^2 g_{\rho_R}^2 = \alpha_{2B}^2 g_{\rho_B}^2\, .
\end{equation}
Under this condition, $\tilde\Pi_1(q^2)/q^2 \to 0$ for $|q^2|\to \infty$, and the integral on the circle vanishes (i.e. $C_1 = 0$ in this model).
The contribution to $\hat S$ from the  tree-level exchange of the resonances, as obtained through the dispersion integral, thus reads
\begin{equation}
\label{eq:Sintoymodel}
\begin{split}
\hat S  & = \frac{g^2}{4}\sin^2\!\theta \, \bigg[ \left( \frac{1}{g_{\rho_L}} -2 \alpha_{2L} g_{\rho_L} \right)^2 + \left( \frac{1}{g_{\rho_R}} -2 \alpha_{2R} g_{\rho_R} \right)^2
                  - 8 \alpha_{2B}^2 g_{\rho_B}^2  \bigg] 
\\[0.2cm]
          & =  \frac{g^2}{4}\sin^2\!\theta \, \bigg[ \left( \frac{1}{g_{\rho_L}^2} -4 \alpha_{2L} \right) + \left( \frac{1}{g_{\rho_R}^2} -4 \alpha_{2R} \right) \bigg] \, ,
\end{split}
\end{equation}
where the second equality follows from Eq.~(\ref{eq:condition}). The expression in the last line coincides with the result of the diagrammatic calculation, 
where the tree-level exchange of the $\rho_B$ gives no contribution to $\hat S$.~\footnote{This can be most easily seen by noticing that integrating out the $\rho_B$
from the Lagrangian~(\ref{eq:LrhoB}) by using the equations of motions does not generate any $O(p^4)$ operator.}
Notice that although $\hat S$ is obtained through a dispersive integral it is not positive definite, because the contribution from the spectral function $\rho_{BB}$
comes with a negative sign in Eq.~(\ref{eq:dispersionS}).

Now consider the limit in which the resonance $\rho_B$ is much heavier than the other two and has a mass $m_{\rho_B} \sim g_* f \gg m_{\rho_L} \sim m_{\rho_R} \sim g_\rho f$.
The scale $m_{\rho_B}$  acts as a cutoff for the effective theory with just $\rho_L$ and $\rho_R$. In such  a low-energy description the leading $O(1/g_\rho^2)$ contribution 
to the $\hat S$ parameter is fully accounted for by the exchange of the light resonances (last line of Eq.~(\ref{eq:Sintoymodel})), and no anomalously large coefficient for 
the  dimension-6 operators  is generated by the cutoff dynamics. The result from the diagrammatic calculation is reproduced by the dispersive approach only after adding
the contribution of the integral on the circle at infinity.
While $\hat S$ is not positive definite, the correct asymptotic behavior of the two-point current correlators is recovered at the cutoff scale through the exchange 
of the $\rho_B$, as a consequence of Eq.~(\ref{eq:condition}). The latter can be satisfied for 
$\alpha_{2L}\sim \alpha_{2R} \sim 1/g_\rho^2$ and $\alpha_{2B} \sim 1/(g_\rho g_*)$.


\end{document}